\documentclass{article}

\usepackage[bottom]{footmisc}
\usepackage{float}
\usepackage{sidecap}
\usepackage{multirow}
\usepackage{parskip}

\usepackage[margin=1.25in]{geometry}
\usepackage[utf8]{inputenc} 
\usepackage[T1]{fontenc}    
\usepackage{url}            
\usepackage{booktabs}       
\usepackage{amsfonts}       
\usepackage{nicefrac}       
\usepackage{microtype}      
\usepackage{kpfonts}
\usepackage{graphicx}
\usepackage{multicol}
\usepackage{enumitem}

\usepackage[small]{caption}
\usepackage{subcaption}

\usepackage[table,dvipsnames]{xcolor}
\definecolor{codegreen}{rgb}{0,0.6,0}
\definecolor{codegray}{rgb}{0.3607843137,
0.4823529412,
0.5725490196}
\definecolor{codeblue}{rgb}{0,0.28,0.67}
\definecolor{backcolour}{rgb}{0.9882352941,
0.9725490196,
0.9294117647}

\usepackage{hyperref}
\hypersetup{
    colorlinks,
    linkcolor={codeblue},
    citecolor={codeblue},
    urlcolor={codeblue}
}

\usepackage[scaled]{beramono}

\usepackage{lipsum}
\usepackage{listings}

\lstdefinestyle{mystyle}{
    escapechar=\%,
    basicstyle=\ttfamily\small,
    breakatwhitespace=false,
    breaklines=true,
    captionpos=b,
    keepspaces=true,
    showspaces=false,
    showstringspaces=false,
    showtabs=false,
    tabsize=2
}
\usepackage{amsmath}
\usepackage{amssymb}
\usepackage{mathtools}
\usepackage{amsthm}

\usepackage{setspace}

\usepackage{adjustbox}
\usepackage{siunitx}
\usepackage{wrapfig}
\usepackage{tcolorbox}
\tcbuselibrary{skins, listings, breakable}
\usepackage{placeins}
\usepackage{algorithmic, algorithm}
\usepackage{svg}
\usepackage{fontawesome}
\usepackage[normalem]{ulem}
\usepackage{textcomp}
\usepackage{xspace}
\usepackage{pifont}
\usepackage{natbib}
\newcommand{\ctick}{\ding{51}}%
\newcommand{\xtick}{\ding{55}}%

\definecolor{burntorange}{RGB}{204,85,0}
\definecolor{uclagold}{RGB}{255, 209, 0}
\definecolor{uclablue}{RGB}{50, 131, 191}
\definecolor{googleblue}{RGB}{66, 133, 244}
\definecolor{googlered}{RGB}{234, 67, 53}
\definecolor{googleyellow}{RGB}{251, 188, 4}
\definecolor{googlegreen}{RGB}{52, 168, 83}

\newcommand{\algoname}{\textsf{LATTICE}}

\newcommand{\todo}[2][]{}

\newcommand{\isd}[1]{}
\newcommand{\nilesh}[1]{}
\newcommand{\cho}[1]{}
\newcommand{\bpngot}[1]{}

\title{\Large
\vspace{-2ex}
LLM-guided Hierarchical Search for End-to-end Reasoning Intensive Retrieval
}
\author{
\small
Nilesh Gupta\thanks{
\,Correspondence to \url{nilesh@cs.utexas.edu}.\\
Code: \url{https://github.com/nilesh2797/llm-guided-hierarchical-search}.\\
}$~\,^{\spadesuit\clubsuit}$,
Wei-Cheng Chang$^{\clubsuit}$,
Ngot Bui$^{\clubsuit}$,
Cho-Jui Hsieh$^{\diamondsuit}$,
Inderjit S. Dhillon$^{\spadesuit\clubsuit}$\\
\small
$^{\spadesuit}$\,UT Austin \quad
$^{\diamondsuit}$\,UCLA \quad
$^{\clubsuit}$\,Google
\vspace{-0.5ex}
}
\date{}

\newcommand{\arxivmode}{}  

\begin{document}

\maketitle

\begin{abstract}
Search systems are increasingly used for \emph{reasoning-intensive} queries, where what makes a document relevant requires reasoning over the query--document relation rather than surface vocabulary or topical similarity. The standard recipe -- a cheap embedding-based retriever followed by an LLM verifier -- works only when the embedding model places the right documents in its top-$k$, an assumption that recent reasoning-intensive benchmarks show often fails even for SOTA embedders~\citep{Su2024-xk,weller2025theoretical}. Query-side fixes such as query rewriting and agentic loops keep the LLM upstream of the cheap retriever and remain brittle to the embedder's failures and to the LLM's ability to rewrite the query from its parametric knowledge. In this paper, we explore a different paradigm -- \emph{LLM-guided hierarchical search} -- in which an LLM interacts with the corpus directly via a hierarchically navigable search index, with no embedding model in the loop at search time. We propose \algoname, an instantiation with two contributions: (i) a top-down LLM-guided construction of the search index using LLM judgements over multi-level document summaries, and (ii) a calibrated, path-aggregated LLM-guided traversal that mitigates noisy, slate-dependent LLM scores via cross-branch reference nodes. On the reasoning-intensive BRIGHT benchmark, base \algoname~with a single off-the-shelf LLM reaches 46.7 nDCG@10 -- matching the best fine-tuned ensemble baseline overall -- and a lightweight ensemble \algoname\textsuperscript{++} that fuses \algoname~with cheap retrieval reaches \textbf{49.1 nDCG@10}. A controlled same-LLM comparison against sliding-window reranking shows reranking wins at low token budgets but \algoname~converges to a higher asymptote after a moderate budget. \algoname~also works with open-weight LLMs and remains competitive on traditional IR benchmarks.

\end{abstract}

\begin{figure}[H]
    \centering
    \begin{minipage}[t]{0.65\linewidth}
        \centering
        \includegraphics[width=\linewidth]{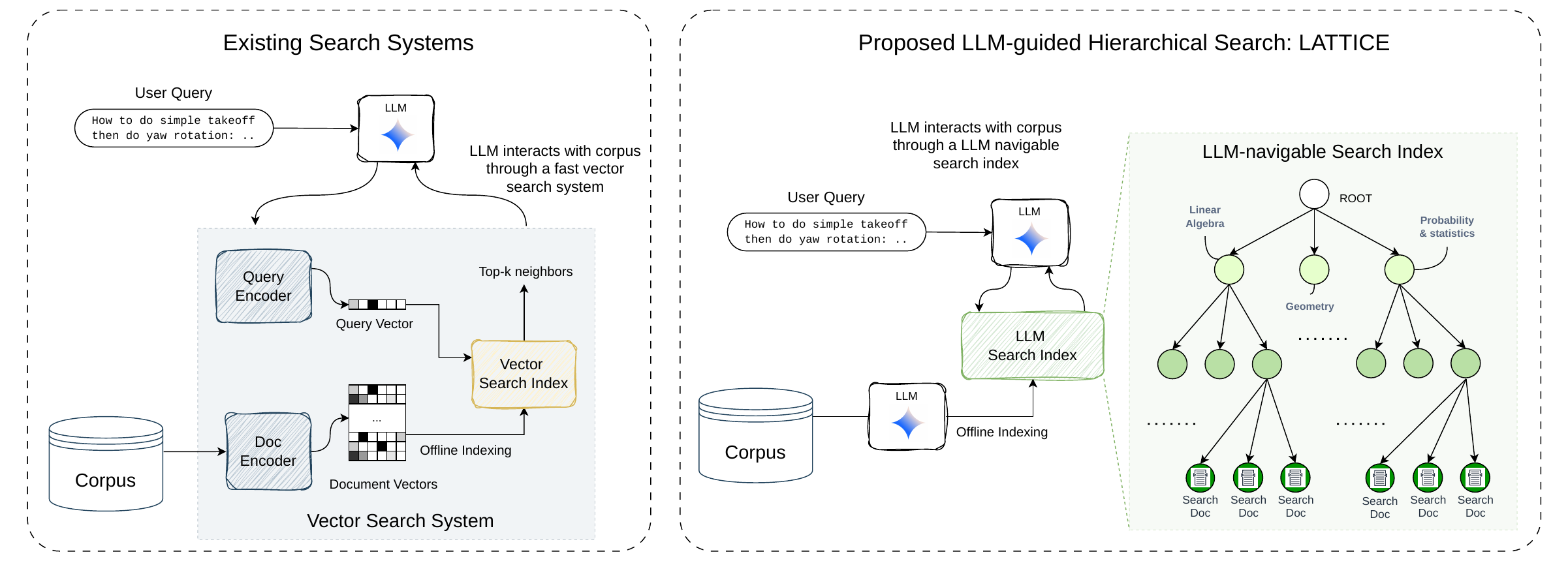}
    \end{minipage}\hspace{0.01\linewidth}%
    \begin{minipage}[t]{0.33\linewidth}
        \centering
        \includegraphics[width=\linewidth]{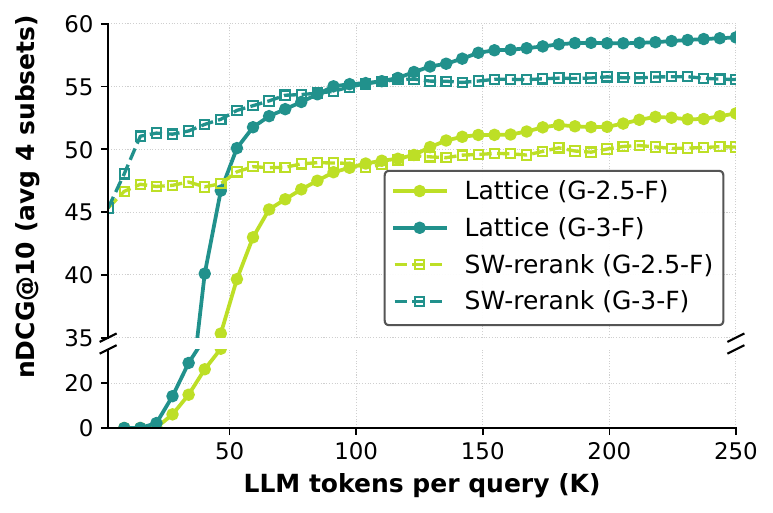}
    \end{minipage}
    \caption{\textbf{\algoname~Overview (left).} Existing systems probe the corpus through a vector search index built using an embedding model. \algoname~replaces the vector index with an \emph{LLM-navigable search index} that an LLM constructs offline from the corpus; at query time, the LLM walks this index directly, with no embedding model in the loop. \textbf{LLM controlled test-time scaling (right).} Averaged over four BRIGHT subsets with Gemini-2.5-Flash and Gemini-3.0-Flash under identical LLM and prompt. Sliding-window reranking (dashed) over a BGE-Reasoner-Embed (with query rewriting) top-300 shortlist wins at low budgets; after a moderate budget \algoname~(solid) overshoots and converges to a higher asymptote.}
    \label{fig:traversal}
    \label{fig:tts_avg_intro}
\end{figure}

\ifdefined\arxivmode\else
\begin{figure*}[!thbp]
    \centering
    \begin{minipage}[t]{0.63\linewidth}
        \centering
        \includegraphics[width=\linewidth]{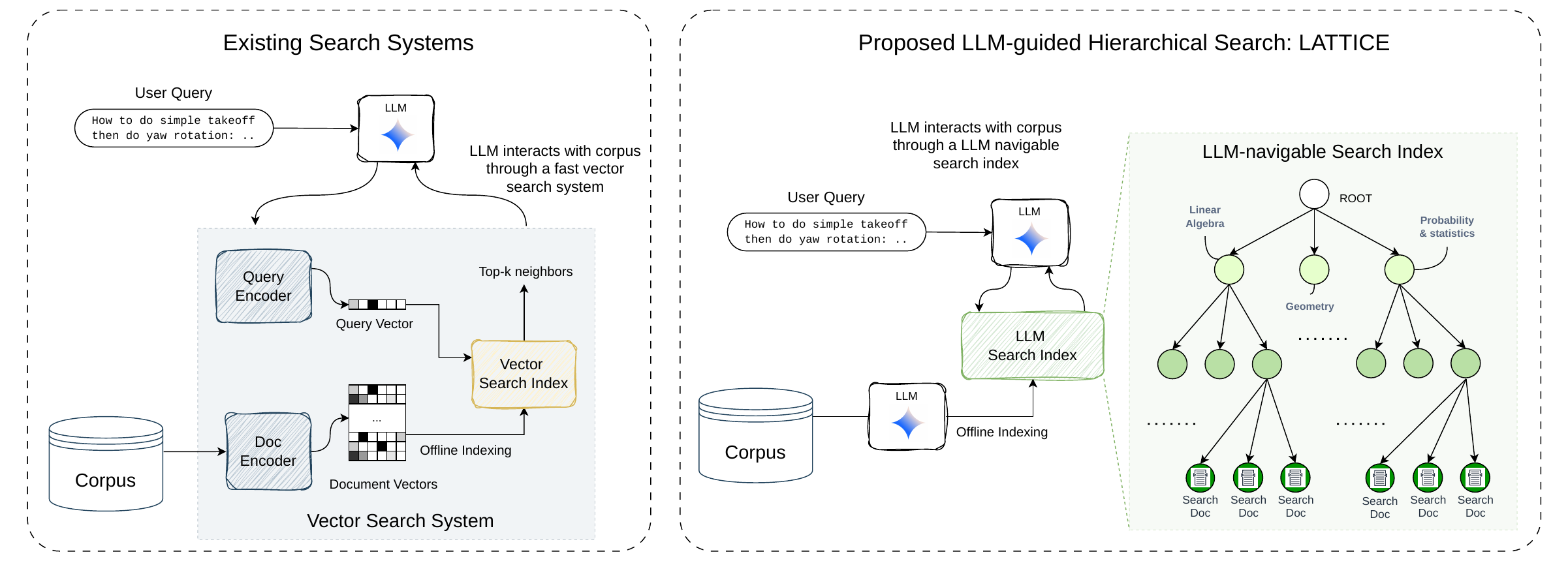}
    \end{minipage}\hspace{0.01\linewidth}%
    \begin{minipage}[t]{0.32\linewidth}
        \centering
        \includegraphics[width=\linewidth]{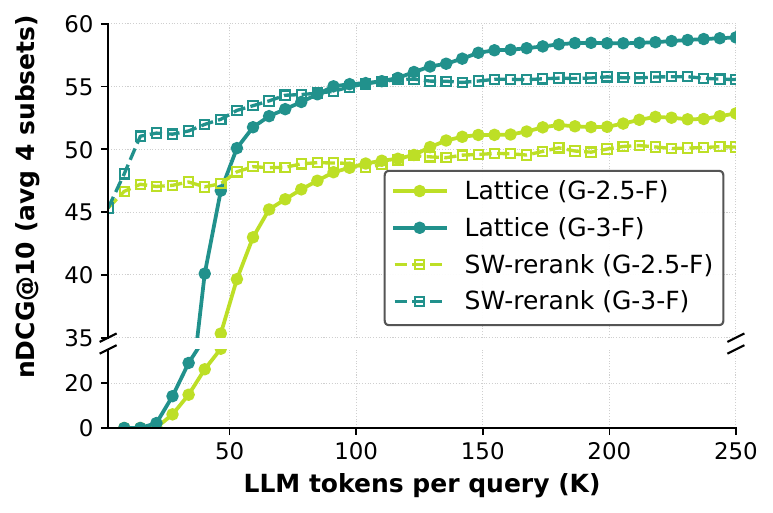}
    \end{minipage}
    \caption{\textbf{\algoname~Overview (left).} Existing systems probe the corpus through a vector search index built using an embedding model. \algoname~replaces the vector index with an \emph{LLM-navigable search index} that an LLM constructs offline from the corpus; at query time, the LLM walks this index directly, with no embedding model in the loop. \textbf{LLM controlled test-time scaling (right).} Averaged over four BRIGHT subsets with Gemini-2.5-Flash and Gemini-3.0-Flash under identical LLM and prompt. Sliding-window reranking (dashed) over a BGE-Reasoner-Embed top-300 shortlist wins at low budgets by inheriting the shortlist's base nDCG@10; after a moderate budget \algoname~(solid) overshoots and converges to a higher asymptote.}
    \label{fig:traversal}
    \label{fig:tts_avg_intro}
    \vspace{-20pt}
\end{figure*}
\fi

\section{Introduction}
\label{sec:intro}

The retrieval workloads that modern search systems are expected to handle are getting harder. A search engine is no longer used only for queries where keyword or topical similarity already pins down the right document. It is increasingly asked to surface evidence for tasks where what makes a document relevant requires reading both the query and the document carefully -- brainstorming a niche research idea, doing deep search on an unfamiliar topic~\citep{he2025pasa}, locating legal precedent for a specific fact pattern~\citep{pipitone2024legalbench}, or searching something with under-specified preferences. In recent literature this is typically referred to as the \emph{reasoning-intensive} retrieval setting~\citep{Su2024-xk,Shao2025-qk}.

The standard information retrieval recipe is to combine a cheap embedding-based retriever (for quick lookup) with a stronger LLM \emph{verifier} (a reranker) that re-orders the retriever's top-$k$~\citep{Sun2023-oa,reddy2024first}. When the embedding model is largely right about which documents belong in the top-$k$, this split is hard to beat -- it is fast, it is cheap, and the verifier handles whatever ordering noise remains. But recent reasoning-intensive benchmarks show that the assumption underneath the split breaks down: even state-of-the-art embedding models fail to place the right documents in the top-$k$ when relevance depends on a relation, an argument, or a chain of inference~\citep{Su2024-xk,Shao2025-qk}, and recent theoretical work shows that no fixed-dimensional dense embedding can represent arbitrary top-$k$ ranking relationships~\citep{weller2025theoretical}. Anything the retriever misses, the verifier cannot recover. The natural question, then, is how to put an LLM's reasoning capacity to work \emph{inside} the retrieval step, not just on top of it.

Most recent work answers this by pushing the LLM upstream of the retriever rather than into it. The cleanest examples are \emph{query rewriting} -- have the LLM produce a hypothetical answer the query might be paired with (HyDE~\citep{gao2023precise}), or expand the user's query into a longer, more informative form (Query2Doc~\citep{wang2023query2doc}) -- and then run a standard retrieve-then-rerank over the rewritten query. \emph{Agentic retrieval} extends this into a loop: read what was retrieved, decide what is still missing, refine the query, retrieve again, repeat until satisfied~\citep{jin2025search,zhang2024agentic}. Both approaches work in many settings, but they share a structural property worth noticing: the strong LLM only indirectly interacts with the corpus. Its reasoning shapes the queries that get sent and the candidates that get re-ordered, but the documents that come back are still chosen by the cheap retriever. This makes the pipeline brittle in several characteristic ways. \textbf{(a)} The rewriting LLM may simply not know beforehand what the right answer looks like~\citep{mallen2023trust}. Without the retriever surfacing genuinely new information at some step, the LLM ends up probing the corpus with variations of its own parametric beliefs, and an agentic loop only reinforces those beliefs rather than correcting them~\citep{xie2024chameleon,sun2025redeep}. \textbf{(b)} Even when the rewrite is good, the embedder may fail to understand a richer, more information-dense query and falls back to surface similarity, surfacing topically related but semantically wrong candidates~\citep{weller2025theoretical}. \textbf{(c)} Many real corpora already come with structure -- Wikipedia articles have an inherent hierarchy of sections and subsections, and long documents are routinely chunked into passages, inducing a natural document-passage hierarchy over the chunkified corpus~\citep{karpukhin2020dpr,lewis2020rag} -- and a flat retrieve-then-rerank pipeline discards this structure entirely, forcing the LLM to score passages out of the context they were written in.

In this paper we propose and explore a new search paradigm -- \emph{LLM-guided hierarchical search} -- in which the LLM interacts with the corpus directly, with no embedding model in the loop at search time. In this paradigm, we first offline build an LLM-navigable hierarchical search index over the corpus: a semantic tree whose leaves are documents and whose internal nodes carry compact LLM-written summaries of the documents below them. At query time, a search LLM walks this index with best-first frontier search, scoring children at selected internal node and descending into the most promising subtree (Figure~\ref{fig:traversal}). Structurally this resembles HNSW-style hierarchical nearest-neighbour search~\citep{malkov2018efficientrobustapproximatenearest}; the difference is that the relevance judgement at each step is produced by an LLM reading content rather than by a vector dot product against a fixed embedding.

Although conceptually simple, making this paradigm work pulls in two non-trivial design problems. \textbf{(1) Building an index the LLM can reliably navigate.} We begin from the natural starting point taken by prior work like RAPTOR~\citep{sarthi2024raptor}: a bottom-up construction driven by an embedding model and an off-the-shelf clustering routine, with an LLM summariser writing each cluster's description. We find that embedding-based clusters give noisy partitions of the corpus -- especially at higher levels, where partitions should reflect broad conceptual distinctions and instead conflate documents that merely share surface vocabulary / topics. We therefore propose to build the hierarchy top-down using LLM judgements over multi-level document summaries, so the same in-context reasoning that makes the LLM a strong ranker is used to organise the corpus it will later have to navigate. \textbf{(2) Robustly navigating it with an off-the-shelf LLM.} An LLM can compare $O(10)$ candidates reliably in a single prompt~\citep{Sun2023-oa,Qin2023-fg,Pradeep2023-ld,zhuang2024setwise,liu2024lostmiddle}, and the scores it returns are noisy and context-dependent -- the same node, presented alongside different siblings, can score very differently~\citep{Sun2023-oa,Qin2023-fg,tang2024foundmiddle,zheng2023judge}. Naively walking the highest-scoring path at each level commits early and abandons branches that turn out to be correct. We address this with two mechanisms: \emph{cross-branch calibration}, in which each scoring prompt is augmented with reference nodes drawn from other parts of the tree so the LLM has a stable comparison baseline across slates; and \emph{path-aggregated scoring}, in which we fit slate-independent latent scores from the full scoring history and accumulate them along the root-to-leaf path, so the search is guided by cumulative evidence rather than by any single local score.

We validate these design choices on the reasoning-intensive BRIGHT benchmark~\citep{Su2024-xk} and three traditional IR benchmarks NQ, SciFact, SciDocs, spanning corpus sizes from 5K to 2.68M documents~\citep{thakur2021beir}. On BRIGHT, base \algoname~with a single zero-shot LLM (Gemini-3-Flash) and no ensembling achieves 46.7 nDCG@10 overall -- exceeding the ensembled BGE-Reasoner-0928 (46.4) by 0.3 and within 0.1 of DIVER~v3 (46.8), and outperforming these fine-tuned ensembles on 5 of the 12 BRIGHT subsets. A lightweight ensemble \algoname\textsuperscript{++} that fuses \algoname~with BM25 and a dense retriever reaches \textbf{49.1 nDCG@10}. In a same-LLM, same-prompt comparison against a sliding-window reranker over a strong dense shortlist (Figure~\ref{fig:tts_avg_intro}), reranking wins at low token budgets but \algoname~converges to a higher asymptote after a moderate budget. \algoname~also works with open-weight LLMs (Qwen3.5-27B matches Gemini-2.5-Flash on StackExchange average at $\sim$4$\times$ lower per-query cost).

Overall, the goal of this paper is to show promise for a new search paradigm in which an LLM interacts with the corpus directly through a navigable search index, rather than probing the corpus through the noisy view of an embedding-based retriever. \algoname~has $O(\log N)$ asymptotic search complexity, but because each step is an LLM call, in its current form it is best suited to applications that can afford a tens of seconds of latency per query -- deep research, legal QA, technical question answering.

\FloatBarrier

\section{Methodology}
\label{sec:method}

This section describes \algoname~in two parts: an offline procedure that organises a corpus into an LLM-navigable hierarchical search index (Section~\ref{sec:construction}), and an online procedure through which a search LLM walks this index to answer a query (Section~\ref{sec:traversal}).

\subsection{Offline Hierarchical Search Index Construction}
\label{sec:construction}

\begin{figure*}[!tbp]
    \centering
    \includegraphics[width=\linewidth]{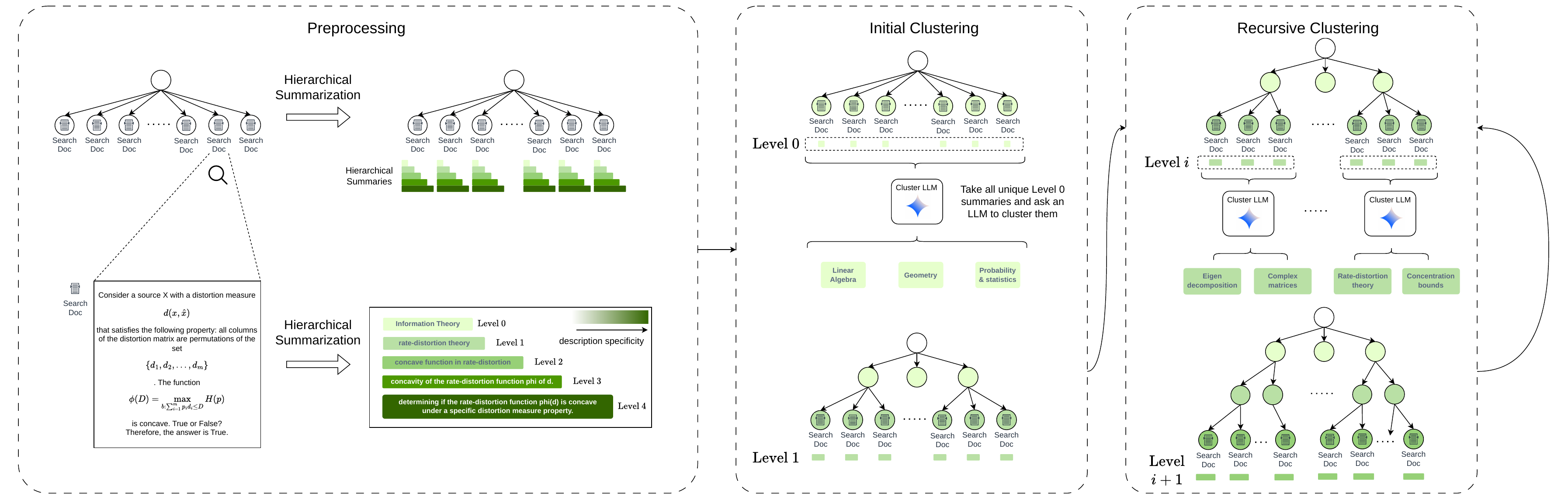}
    \caption{Top-down construction of the \algoname~index. Starting from a root containing all leaf documents, we recursively partition any node with more than $M$ children: an LLM is shown multi-level summaries of the node's leaves and asked to organise them into at most $M$ coherent topics. Each topic becomes a new internal node whose LLM-written description serves as $\phi(v)$; leaves are reassigned per the LLM's mapping.}
    \label{fig:tree_construction}
    \vspace{-10pt}
\end{figure*}

We formalise the index as a rooted tree $T = (V, E, \phi)$. The node set $V = V_L \cup V_I$ consists of leaf nodes $V_L$ -- one per document in the corpus $D$ -- and internal nodes $V_I$, each representing a cluster of leaf documents below it. Every node carries a textual representation $\phi(v)$: for leaves, the document content; for internal nodes, an LLM-written summary of the documents in its subtree. We write $C(v)$ for the children of $v$ and constrain the branching factor to $|C(v)| \le M$.

\paragraph{Bottom-up construction.}
To build such an index, we begin from the natural starting point taken by prior work like RAPTOR~\citep{sarthi2024raptor}: a bottom-up construction driven by an embedding model $\mathcal{E}$ and an off-the-shelf clustering routine such as $k$-means. Concretely, starting from $V_0 = V_L$ (one node per document), at each level $\ell = 0, 1, \ldots$, we embed the textual representations $\{\mathcal{E}(\phi(v)) : v \in V_\ell\}$, cluster the resulting vectors into at most $M$ groups, instantiate a new internal node per group whose children are the corresponding $V_\ell$ nodes, and use a summariser LLM to write the new node's $\phi(\cdot)$ from the textual representations of its children. The procedure terminates once $|V_\ell| \le M$, at which point the surviving nodes become the children of a single root. The full procedure is given as Algorithm~\ref{alg:construction} in Appendix~\ref{app:clustering}.

While conceptually simple and practical to implement, we find that bottom-up indices do not work well at the corpus sizes considered in this work, particularly when passages come from independent sources rather than from a shared parent document; Table~\ref{tab:ablation_tree} reports the corresponding ablation. We hypothesise that this is because a RAPTOR-style index inherits the noise behaviour of the embedding model on which it is built. At the lowest levels of the tree the noise is benign -- local neighbourhoods of nearly-synonymous documents are well-formed by embedding similarity, since at this granularity surface similarity and conceptual similarity do tend to agree. At higher levels, however, the partition should reflect broad conceptual distinctions (e.g.\ separating \emph{population dynamics} from \emph{biochemical signalling} in a biology corpus), and embedding-based clusters tend to mix such conceptually distinct sub-clusters whenever they happen to share surface vocabulary~\citep{weller2025theoretical}. The noisy higher-level partitions are then summarised by an LLM, locking the conflation into the cluster summaries the search will later see.

\paragraph{Top-down construction with an LLM.}
To address this, we propose \algoname's \emph{top-down tree construction} algorithm: a fully LLM-supervised clustering procedure that replaces the embed-and-cluster step at every level with an LLM call. The high-level structure mirrors hierarchical $k$-means: begin with a single root containing every leaf, and recursively partition any cluster whose size exceeds a threshold. The key change is operationalisation -- we want the LLM, rather than an embedding clustering routine, to decide how to partition each cluster~\citep{viswanathan2024llmclustering,pham2024topicgpt}.

The simplest LLM-only instantiation is impractical. Given a node $v$ to partition, we would concatenate the textual representations $\{\phi(v_i) : v_i \in V_L(v)\}$ of all leaves under $v$ and ask an LLM to (i) identify at most $M$ topical clusters with short titles and (ii) map each input to one cluster. Even for a moderately sized cluster ($\sim 10^4$ documents at a few hundred words each), the prompt is already in the millions of tokens, well outside the context limits of any LLM we can afford to call recursively along the tree.

\paragraph{Hierarchical summarisation pre-processing.}
To make the LLM-only construction practical, we preprocess each document with a hierarchical summarisation step (left panel of Figure~\ref{fig:tree_construction}). A lightweight summariser LLM $\mathcal{S}$ produces $k$ summaries of each leaf's textual representation $\phi(v_l)$ at increasing specificity, with the rule that the level-$i$ summary uses at most $2^{i+1}$ words, so the descriptive capacity doubles at every level. We write the resulting summaries as
\[
\phi^{(i)}(v_l) \;=\; \mathcal{S}\bigl(\phi(v_l),\, i\bigr), \qquad i = 0, 1, \ldots, k-1, \qquad |\phi^{(i)}(v_l)|_{\text{words}} \le 2^{i+1}.
\]
The summaries trade off context length for descriptive specificity: $\phi^{(0)}(v_l)$ is a one-to-two-word topic and $\phi^{(k-1)}(v_l)$ is a one-sentence description (a worked example for a rate-distortion-theory document is shown in Figure~\ref{fig:tree_construction}; the full summariser prompt is in Figure~\ref{fig:multi_level_keyword_prompt}, Appendix~\ref{app:prompts}). We use $k = 5$ throughout, which is roughly $\log_M N$ deep for the corpora we work with ($N \in [10^4, 10^6]$, $M \in [10, 20]$). For efficiency, $\mathcal{S}$ is realised by a small LLM (Gemini-3.1-Flash-Lite or Qwen3-4B); to amortise the prompt overhead we batch the summarisation, feeding 20 documents in one prompt and asking the model to return hierarchical summaries for all 20 in its output.

\paragraph{Hierarchical summaries as a dynamic interface to the corpus.}
Once the multi-level summaries are available, they serve as a token-efficient, depth-adaptive interface between a cluster LLM $\textsc{ClusterLLM}$ and the (sub-)corpus it is asked to partition. To split a cluster node $v$ into smaller clusters, we proceed as follows (Algorithm~\ref{alg:top_down_construction}; the middle and right panels of Figure~\ref{fig:tree_construction} illustrate one application at the root and one recursive application deeper in the tree):
\begin{enumerate}[leftmargin=*, itemsep=2pt, topsep=2pt]
\item \textbf{Choose the summary level.} Let $V_L(v)$ be the leaves under $v$. A \textsc{SelectSummaryLevel} subroutine picks the most specific level $i^\star$ such that the deduplicated set $U^{(i^\star)}(v) = \{\phi^{(i^\star)}(v_l) : v_l \in V_L(v)\}$ fits within $\textsc{ClusterLLM}$'s context budget. This adapts to the depth of the partition: near the root the level is coarse (a few-word topic per document), deeper in the tree it is fine (a sentence per document).
\item \textbf{Cluster.} The deduplicated set $U^{(i^\star)}(v)$ is passed to $\textsc{ClusterLLM}$ (prompt in Figure~\ref{fig:cluster_llm_prompt}, Appendix~\ref{app:prompts}), which is asked to (i) identify between $M_{\min}$ and $M$ topical clusters along with a short LLM-written title per cluster, and (ii) return a mapping from each input summary to a cluster.
\item \textbf{Reassign.} Each topical cluster becomes a new internal node, with the cluster title as its representation $\phi(\cdot)$. Every leaf $v_l \in V_L(v)$ is reassigned to the new internal node corresponding to $\phi^{(i^\star)}(v_l)$ under the mapping.
\item \textbf{Recurse.} Any newly created internal node whose child count exceeds $M$ is added to a partitioning queue and processed in the same way. The recursion terminates when every node satisfies $|C(v)| \le M$.
\end{enumerate}
The depth-adaptive level choice in step 1 means partitions near the root are made over coarse summaries -- enough to discriminate broad domains -- while partitions deeper in the tree are made over fine summaries that capture finer topical distinctions.

\paragraph{Hybrid construction for large corpora and corpora with a natural parent--child hierarchy.}
In practice, two situations call for a hybrid construction that combines a bottom-up first pass with the proposed top-down algorithm. (a)~When the corpus already carries a natural parent--child hierarchy -- e.g.\ chunks derived from longer documents such as Wikipedia articles or research papers -- we use that hierarchy directly to form the initial clusters (each parent document becomes a cluster whose leaves are its chunks). (b)~When the corpus is too large for the top-down clustering to run directly over all $N$ leaves (as is the case for NQ at $2.6$M passages), we first run the bottom-up construction of Algorithm~\ref{alg:construction} to obtain an initial layer of coarse clusters, and treat those as pseudo-leaves. In both cases, the top-down algorithm above is then applied with the resulting pseudo-leaves in place of $V_L$. This is more efficient as the LLM-driven top-down clustering operates on a much smaller leaf set -- and, in case (a), it also preserves the natural organisation of the corpus inside the index, so that the search LLM can later evaluate individual passages within the context of their parent document.

\subsection{Online LLM-guided Hierarchical Search}
\label{sec:traversal}

\begin{figure*}[!tbp]
    \centering
    \includegraphics[width=\linewidth]{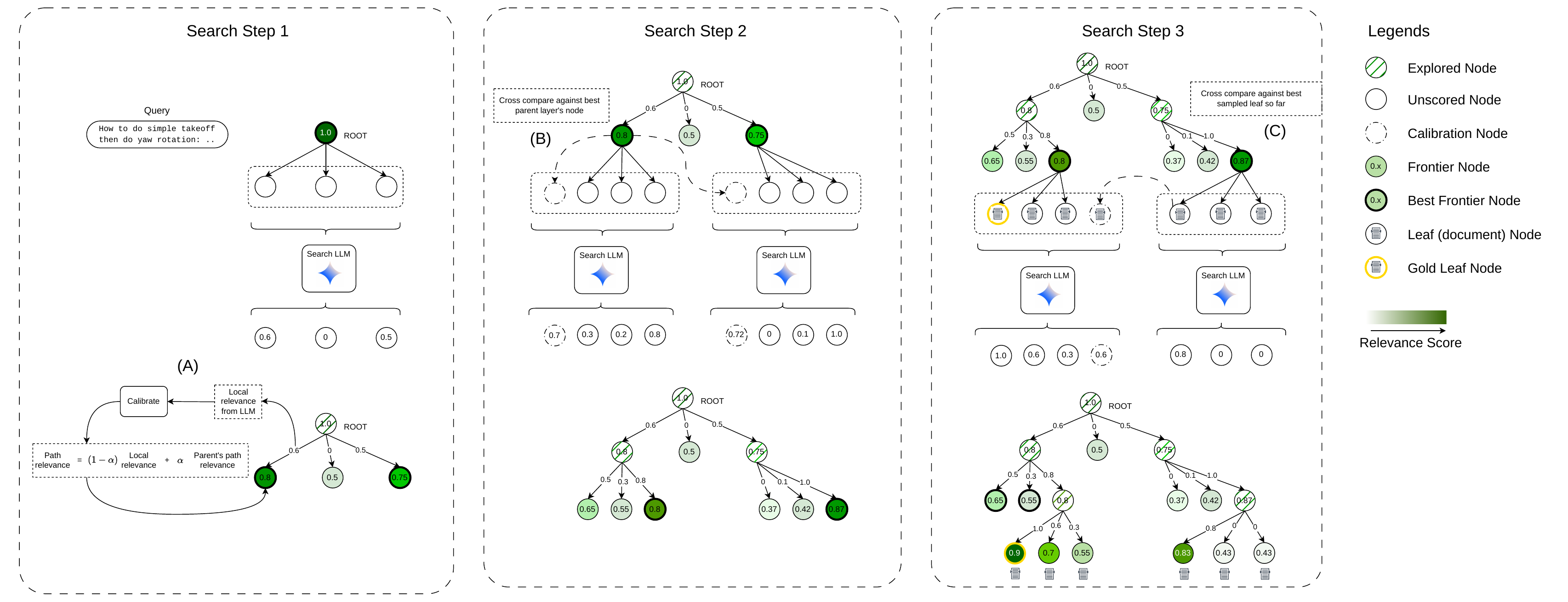}
    \caption{\algoname~tree search. A frontier $F$ of unexpanded internal nodes is ordered by path-relevance. At each iteration the top-$B$ frontier nodes are expanded: the LLM scores a slate of each node's children plus cross-branch reference nodes; raw scores are converted to slate-independent latent scores by fitting a per-slate-bias linear model over the running history; latent scores are accumulated along the root-to-leaf path into a global path-relevance score that drives the next iteration's frontier priority.}
    \label{fig:tree_search}
    \vspace{-10pt}
\end{figure*}

At query time, the index from Section~\ref{sec:construction} is walked by an LLM-guided best-first frontier search. At each search step we use the LLM as an in-context \emph{listwise scorer}: given a query $q$ and the textual descriptions of a small set of candidate nodes, it returns a graded relevance score in $[0, 1]$ per candidate. Formally, the search LLM is a function
\begin{equation}
\mathcal{L}(q,\,[\phi(v_1),\ldots,\phi(v_n)]) \;=\; [s_1,\ldots,s_n], \qquad s_i \in [0,1].
\label{eq:scorer}
\end{equation}
The prompt has four blocks -- instruction, query, candidate descriptions, and a JSON output schema requesting per-candidate reasoning and a score in $[0,1]$; the full template is in Figure~\ref{fig:scoring_prompt}, Appendix~\ref{app:prompts}. Listwise scoring lets the LLM compare candidates in context within a single call and reduces the per-expansion budget from $n$ pointwise calls to one.

It is useful to read each per-child score $s_i$ as the LLM's estimate of the \emph{value} of expanding the subtree rooted at $v_i$ -- loosely, how relevant the best leaf reachable under $v_i$ is to the query. The search then behaves like a best-first heuristic search whose value function is an LLM reading content~\citep{hart1968astar,russell2021aima}; this LLM-as-value-function framing has also seen recent use in LLM tree-search for reasoning~\citep{yao2023tot,hao2023rap}. The non-trivial part is converting these per-step value estimates into a globally consistent ranking over the tree. A naive instantiation -- score each frontier node's children with $\mathcal{L}$, push them onto the frontier by raw score, repeat -- fails for two reasons:
\begin{itemize}[leftmargin=*, itemsep=2pt, topsep=2pt]
\item \textbf{Slate dependence.} A listwise scorer's output depends on which other candidates appear in the prompt~\citep{Sun2023-oa,Qin2023-fg}. A node $v$ scored alongside a strong sibling can receive a lower number than the same $v$ scored alongside a weak one; reading raw scores as direct estimates of $v$'s relevance conflates the LLM's view of $v$ with the company $v$ happened to keep.
\item \textbf{Locality.} Even a perfectly comparable raw score reflects a comparison among only the $\le M$ candidates in a single slate, which is too narrow a signal to globally order the thousands of paths the search is choosing between.
\end{itemize}
Our expansion routine addresses both: a \emph{latent-score estimator} strips the slate-dependent bias from the raw $s^i_v$, and a \emph{path-relevance} aggregator carries information along the search trajectory. The rest of this section describes the search loop and one full expansion in execution order; the full pseudocode is in Algorithm~\ref{alg:traversal} (Appendix~\ref{app:implementation}).

\paragraph{Search loop.}
The search (Figure~\ref{fig:tree_search}; pseudocode in Algorithm~\ref{alg:traversal}, Appendix~\ref{app:implementation}) maintains
\begin{itemize}[leftmargin=*, itemsep=2pt, topsep=2pt]
\item a max-priority queue $F$ of unexpanded internal nodes, keyed by a path-relevance score $\hat{p}_{\text{rel}}(v)$ defined below;
\item a running prediction set $\text{Pred}$ of leaves already scored;
\item a score history of $(\text{slate id},\, v,\, s^i_v)$ triples accumulated from every listwise call.
\end{itemize}
We seed $F$ with the root and set $\hat{p}_{\text{rel}}(v_{\text{root}}) = 1$. Each iteration pops the top-$B$ entries of $F$ and expands them in parallel as described in the next four paragraphs. After $N$ iterations the search returns the top-$K$ entries of $\text{Pred}$ ranked by $\hat{p}_{\text{rel}}$.

\paragraph{Slate construction.}
For each expanded node $v$, the LLM is shown a \emph{slate} -- the list of candidates to score in a single listwise call. The slate is $v$'s children $C(v)$ together with a small calibration set $\text{Aug}(v)$ drawn from elsewhere in the tree. The calibration set has a single purpose: it gives the LLM reference candidates that recur across slates, anchoring the resulting scores to a strong comparison. The composition of $\text{Aug}(v)$ depends on the type of children $v$ has:
\begin{itemize}[leftmargin=*, itemsep=2pt, topsep=2pt]
\item \textbf{When $C(v)$ are internal nodes:} $\text{Aug}(v)$ contains the highest-scoring sibling of $v$ from a previously expanded slate -- a cross-branch anchor the LLM has already evaluated.
\item \textbf{When $C(v)$ are leaves:} $\text{Aug}(v)$ contains $\ell$ top-scoring leaves drawn from $\text{Pred}$, sampled with probability proportional to their current latent score -- the strongest leaves seen so far, used as anchors.
\end{itemize}
The full slate is passed to $\mathcal{L}$ (Eq.~\ref{eq:scorer}), which returns raw scores $s^i_{v'} \in [0,1]$ for every item in slate $i$. The resulting triples $(i, v', s^i_{v'})$ are appended to the score history.

\paragraph{Latent score estimation.}
Calibration anchors do not by themselves eliminate slate dependence -- a node's raw score still varies with the particular composition of its slate. We estimate a slate-independent \emph{latent score} $\hat{s}_v$ per node by modelling each observed raw score as a linear function of the latent score, plus a slate-specific shift -- in the same spirit as recent work that calibrates noisy LLM-judge scores by fitting a latent-skill model over many comparisons~\citep{zhou2024calibraeval,zheng2023judge}:
\begin{equation}
s^i_v \;\approx\; a \cdot \hat{s}_v \;+\; b^i,
\label{eq:latent}
\end{equation}
where $a$ is a global scale shared across slates and $b^i$ is a per-slate bias. After every expansion, all parameters are re-estimated jointly by minimising mean-squared error over the full score history,
\begin{equation}
\min_{a,\,\{\hat{s}_v\},\,\{b^i\}} \;\; \sum_i \sum_{v' \in \text{slate}^i} \bigl(s^i_{v'} - (a \cdot \hat{s}_{v'} + b^i)\bigr)^2.
\label{eq:latent_mle}
\end{equation}
A node that appears across multiple slates has its $\hat{s}_v$ jointly constrained by all of them. The per-slate $b^i$ term is what distinguishes the estimator from a naive per-node average over raw scores: averaging conflates a node's intrinsic relevance with the strength of the company it happened to keep, biasing estimates downward for nodes that tended to appear alongside stronger candidates and upward for those that tended to appear alongside weaker ones.

\paragraph{Path-relevance update.}
Latent scores are still local -- each $\hat{s}_v$ reflects a comparison among the $O(M)$ candidates in $v$'s slate, a narrow signal for a search over $\Theta(N)$ documents. To carry information along the search trajectory, we accumulate latent scores along the root-to-node path into a \emph{path-relevance} score
\begin{equation}
\hat{p}_{\text{rel}}(v) \;=\; \alpha \cdot \hat{p}_{\text{rel}}(\text{parent}(v)) \;+\; (1-\alpha) \cdot \hat{s}_v,
\label{eq:pathrel}
\end{equation}
with $\hat{p}_{\text{rel}}(v_{\text{root}}) = 1$ and $\alpha \in [0, 1]$ controlling the trade-off between ancestral and local evidence. This is an exponential moving average up the path; with $\alpha = 0.5$ (our default) it weights deeper (more recent) splits more heavily, which we want because they are made over finer-grained summaries that more directly reflect the leaves below.

\paragraph{Frontier and prediction-set update.}
After expanding $v$, each newly scored internal child $v' \in C(v)$ is pushed onto $F$ with priority $\hat{p}_{\text{rel}}(v')$, and each leaf child is added to $\text{Pred}$. The next iteration pops the top-$B$ entries of $F$ and repeats. The search terminates after $N$ iterations and returns the top-$K$ entries of $\text{Pred}$ ranked by $\hat{p}_{\text{rel}}$, for a per-query cost of $B \cdot N$ listwise LLM calls and asymptotic $O(\log_M N)$ tree depth.

\section{Experiments}
\label{sec:experiments}

This section evaluates \algoname~along three axes: ranking quality on the reasoning-intensive BRIGHT benchmark (Section~\ref{sec:results}); scalability to traditional IR corpora up to 2.68M documents (Section~\ref{sec:beir}); and per-component contribution via ablations (Section~\ref{sec:ablations}).

\subsection{Experimental Setup}
\label{sec:setup_exp}

\paragraph{Benchmark and metrics.}
BRIGHT~\citep{Su2024-xk} contains 12 reasoning-intensive retrieval tasks (corpus sizes 7.9K--414K) where queries require multi-step inference rather than keyword match. Three subsets (marked $^*$ in Table~\ref{tab:ndcg10_bright}) use a \emph{query-dependent dynamic corpus} -- up to 10K documents are excluded per query at eval time -- which we discuss in Section~\ref{sec:dynamic_corpus}. We report nDCG@10 for ranking quality and Recall@100 for coverage. Per-dataset statistics are in Appendix~\ref{app:datasets}.

\paragraph{Baselines.}
We compare against the strongest published systems on BRIGHT: ReasonIR~\citep{Shao2025-qk}, DIVER~v1/v2/v3~\citep{long2025diver}, ReasonRank~\citep{liu2025reasonrank}, BGE-Reasoner~\citep{chen2025reasonembed}, and XRR2. The current leaders -- DIVER and BGE-Reasoner -- are fine-tuned ensembles combining four retrieval sources (dense retriever, BM25, pointwise reranker, listwise reranker)~\citep{long2025diver,chen2025reasonembed}.

\paragraph{Implementation.}
We use Gemini-2.5-Flash as the summariser LLM and the \textsc{ClusterLLM} for offline index construction. For the main comparison we test two search LLMs -- Gemini-2.5-Flash and the stronger Gemini-3-Flash -- to measure search-LLM-capability effects. Ablations use Gemini-3-Flash unless stated otherwise. Default hyperparameters: $N{=}20$ iterations, beam $B{=}2$, path-relevance momentum $\alpha{=}0.5$, $\ell{=}10$ leaf calibration nodes, branching factor $M{=}10$--$20$. Index construction for a 100K-document corpus (e.g.\ Biology) takes $\approx$6 hours; more details in Appendix~\ref{app:implementation}.

\paragraph{\algoname\textsuperscript{++} ensemble.}
For a better comparison against the fine-tuned ensemble leaders on the BRIGHT leaderboard -- which combine multiple retrieval sources -- we report a lightweight ensemble \algoname\textsuperscript{++} that augments \algoname~with two cheap external signals. For each query we take per-document scores from three sources (\algoname, BM25, and the BGE-Reasoner-Embed dense retriever), min-max normalise each source's scores to $[0,1]$, and form the per-document fused score as a weighted sum with weights $w_{\algoname}{=}0.6$, $w_{\text{BM25}}{=}0.2$, $w_{\text{BGE-RE}}{=}0.2$. BM25 and BGE-Reasoner-Embed are both run on the BGE-reasoner rewritten query. A per-component decomposition of the ensemble is in Appendix~\ref{app:ensemble_ablation}.

\subsection{Main Results on BRIGHT}
\label{sec:results}

\begin{table*}[!tbp]
    \caption{nDCG@10 performance of various retrievers and rankers on the BRIGHT benchmark. \textbf{Bold} represents overall best numbers, \underline{underline} represents best numbers among non-ensemble methods, $^*$ denotes subsets with dynamic corpus.}
    \label{tab:ndcg10_bright}
    \begin{center}
    \begin{small}
    \resizebox{\textwidth}{!}{%
    \begin{tabular}{l|c|c|ccccccc|c|cc|c|ccc|c}
    \toprule
    \textbf{Method} & \textbf{Ensem-} & \multicolumn{8}{c}{\textbf{StackExchange}} & \multicolumn{3}{c}{\textbf{Coding}} & \multicolumn{4}{c}{\textbf{Theorem-based}} & \textbf{Avg.} \\
    \cmidrule(lr){3-10} \cmidrule(lr){11-13} \cmidrule(lr){14-17}
    & \textbf{ble} & \textbf{Avg.} & \textbf{Bio.} & \textbf{Earth.} & \textbf{Econ.} & \textbf{Psy.} & \textbf{Rob.} & \textbf{Stack.} & \textbf{Sus.} & \textbf{Avg.} & \textbf{Leet.$^*$} & \textbf{Pony} & \textbf{Avg.} & \textbf{AoPS$^*$} & \textbf{ThQ.$^*$} & \textbf{ThT.} \\
    \midrule
    \multicolumn{18}{c}{Retrieve-then-rerank} \\
    \midrule
    Rank-R1-14B               & \xtick & 26.6 & 31.2 & 38.5 & 21.2 & 26.4 & 22.6 & 18.9 & 27.5 & 14.7 & 9.2  & 20.2 & 10.3 & 9.7  & 11.9 & 9.2  & 20.5 \\
    Qwen1.5-7B + InteRank-3B  & \xtick & 31.9 & 51.2 & 51.4 & 22.4 & 31.9 & 17.3 & 26.6 & 22.4 & 23.8 & 24.5 & 23.1 & 19.4 & 13.5 & 19.3 & 25.5 & 27.4 \\
    GPT-4 + Rank1-32B         & \xtick & 32.0 & 49.7 & 35.8 & 22.0 & 37.5 & 22.5 & 21.7 & 35.0 & 25.7 & 18.8 & 32.5 & 25.8 & 10.8 & 22.9 & 43.7 & 29.4 \\
    ReasonIR & \xtick & 41.7 & 59.8 & 53.2 & 32.0 & 43.6 & 28.8 & 38.7 & 36.0 & 34.0 & 33.2 & 34.8 & 29.4 & 7.9 & 32.6 & 47.7 & 37.3 \\
    ReasonRank & \xtick & 46.8 & 62.7 & 55.5 & 36.7 & 54.6 & 35.7 & 38.0 & 44.8 & 27.5 & 29.5 & 25.6 & 35.5 & 14.4 & 42.0 & 50.1 & 40.8 \\
    XRR2 (GPT-4+BM25 $\rightarrow$ G-2.5-F) & \xtick & 47.4 & 63.1 & 58.2 & 38.5 & 52.9 & 37.1 & 37.6 & 44.6 & 28.4 & 21.9 & 35.0 & 31.8 & \underline{15.7} & 34.4 & 45.5 & 40.3 \\
    \midrule
    \multicolumn{18}{c}{Iterative-retrieve-then-rerank} \\
    \midrule
    BGE-Reasoner (w/o ensembling) & \xtick & 51.1 & 64.3 & 62.6 & 45.3 & 52.7 & 44.3 & 41.5 & 46.9 & 22.3 & 14.4 & 30.2 & 34.5 & 14.2 & 41.7 & 47.5 & 42.2 \\
    DIVER v2 & \ctick & 52.2 & 68.0 & 62.5 & 42.0 & 58.2 & 41.5 & 44.3 & 49.2 & 33.8 & 34.8 & 32.9 & 38.6 & \textbf{19.1} & 44.3 & 52.6 & 45.7 \\
    BGE-Reasoner-0928 & \ctick & 52.0 & 68.5 & 66.4 & 40.6 & 53.1 & 43.2 & 44.1 & 47.8 & 35.3 & 29.0 & 41.6 & \textbf{40.7} & 17.2 & 46.5 & \textbf{58.4} & 46.4 \\
    DIVER v3 & \ctick & 51.8 & 66.0 & 63.7 & 42.4 & 55.0 & 40.6 & 44.7 & \textbf{50.4} & 39.9 & 32.5 & 47.3 & 39.7 & 17.2 & 46.4 & 55.6 & 46.8 \\
    \midrule
    \multicolumn{18}{c}{LLM-guided Hierarchical Search} \\
    \midrule
    \algoname~(G-2.5-f)     & \xtick & 52.0 & 66.3 & 63.0 & 47.4 & 54.0 & 47.6 & 37.6 & 48.2 & 26.9 & 19.9 & 34.0 & 32.6 & 12.0 & 38.0 & \underline{47.9} & 43.0 \\
    \algoname~(G-3.0-f)     & \xtick & \textbf{\underline{55.8}} & \underline{69.4} & \underline{65.0} & \textbf{\underline{48.8}} & \textbf{\underline{61.0}} & \textbf{\underline{53.9}} & \underline{42.2} & \textbf{\underline{50.4}} & \underline{33.0} & \underline{22.9} & \underline{43.2} & \underline{34.6} & 13.7 & \underline{44.0} & 46.0 & \underline{46.7} \\
    \algoname\textsuperscript{++} (G-2.5-f) & \ctick & 53.6 & 71.4 & 66.9 & 46.5 & 55.3 & 45.4 & 43.7 & 45.9 & 36.7 & 28.5 & 44.8 & 36.7 & 14.2 & 43.6 & 52.4 & 46.5 \\
    \algoname\textsuperscript{++} (G-3.0-f) & \ctick & 55.7 & \textbf{73.2} & \textbf{67.1} & 45.3 & 58.3 & 52.4 & \textbf{45.5} & 48.2 & \textbf{41.9} & 32.6 & \textbf{51.2} & 38.3 & 16.2 & \textbf{46.9} & 51.9 & \textbf{49.1} \\
    \bottomrule
    \end{tabular}%
    }
    \end{small}
    \end{center}
    \vskip -0.1in
\end{table*}

\paragraph{vs.\ state-of-the-art systems.}
With Gemini-3-Flash as the search LLM, base \algoname~reaches 46.7 nDCG@10 overall -- exceeding the four-component fine-tuned ensemble BGE-Reasoner-0928 (46.4) by 0.3 and trailing DIVER~v3 (46.8) by 0.1, while remaining single-LLM, zero-shot, and fine-tuning-free. With the weaker Gemini-2.5-Flash, \algoname~reaches 43.0, behind the best ensembles by 3--4 points. Among non-ensemble baselines, \algoname~with Gemini-3-Flash beats the strongest (BGE-Reasoner without ensembling, 42.2) by $+$4.5 points overall and the next-best zero-shot baseline XRR2 (40.3) by $+$6.4, with a $+$8.4 gap on StackExchange average (55.8 vs.\ 47.4). The low-cost ensemble \algoname\textsuperscript{++} (Section~\ref{sec:setup_exp}) pushes this further: with Gemini-3-Flash it reaches \textbf{49.1 nDCG@10} overall, ahead of DIVER~v3 (46.8) and BGE-Reasoner-0928 (46.4) by 2.3--2.7 points, and setting a new best on 8 of 16 columns in Table~\ref{tab:ndcg10_bright}. The gain over base \algoname~is largest on the dynamic-corpus subsets where the static index suffers from stale summaries (Section~\ref{sec:dynamic_corpus}).

\begin{figure}[!t]
    \centering
    \includegraphics[width=\linewidth]{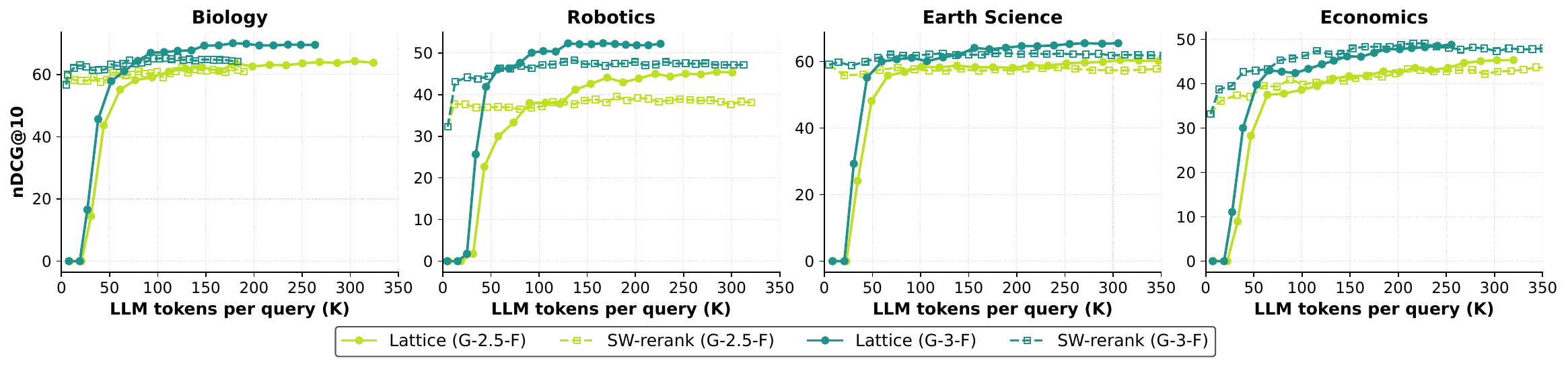}
    \vspace{-12pt}
    \caption{Same-LLM test-time scaling on four BRIGHT subsets. Solid = \algoname; dashed = sliding-window reranker over a BGE-Reasoner-Embed top-300 shortlist; yellow = Gemini-2.5-Flash, teal = Gemini-3-Flash. Reranking starts ahead by inheriting the shortlist's base nDCG@10 at zero rerank cost, then plateaus once its retrieval ceiling is hit; \algoname~keeps improving with deeper exploration.}
    \label{fig:cost_performance}
\end{figure}

\paragraph{Controlled comparison vs.\ retrieve-then-rerank.}
The cleanest measure of the algorithmic gap between \algoname~and retrieve-then-rerank is a same-LLM, same-prompt comparison. We use a sliding-window reranker (SW-rerank) that takes the BGE-Reasoner-Embed top-300 shortlist over the BGE-Reasoner rewritten query and reranks it with the \emph{same} search LLM and the \emph{same} scoring prompt as \algoname; only the search algorithm differs. Following standard practice, SW-rerank uses a window size of 20 with a top-10 carry-over between successive windows. Figure~\ref{fig:cost_performance} reports per-subset token--accuracy curves on Biology, Robotics, Earth Science, and Economics. SW-rerank wins at low token budgets because the shortlist itself already has non-trivial nDCG@10 essentially for free (just the query-rewriting cost), but it plateaus once its retrieval ceiling is hit; \algoname~overshoots SW-rerank after a moderate budget and converges to a higher asymptote on Biology ($+5.3$ with Gemini-3-Flash), Robotics ($+5.0$), and Earth Science ($+3.8$), and matches it on Economics ($+0.7$). The crossover holds for both Gemini-2.5-Flash and Gemini-3-Flash, with the gap widening on the stronger LLM.

\begin{table}[!h]
\centering
\small
\caption{nDCG@10 with open-source and proprietary search LLMs on five BRIGHT subsets (zero-shot, $B{=}2$, $N{=}20$, $\ell{=}10$). \textbf{Cost per query} is computed from input/output tokens measured directly from \algoname~runs and per-token rates from Google AI Studio (Gemini) and DeepInfra (Qwen); full pricing and token counts are listed in Appendix~\ref{app:cost_breakdown}. Qwen and Gemini-3.1-Flash-Lite run in non-thinking mode; Gemini-2.5-Flash and Gemini-3-Flash use thinking.}
\label{tab:open_source_llms}
\begin{tabular}{lccccc|cc}
\toprule
\textbf{Model} & \textbf{Bio} & \textbf{Ear} & \textbf{Eco} & \textbf{Psy} & \textbf{Rob} & \textbf{Avg} & \textbf{Cost / query} \\
\midrule
Qwen3.5-0.8B   & 2.9  & 2.5  & 0.3  & 1.5  & 0.9  & 1.6  & \$0.003 \\
Qwen3.5-2B     & 16.0 & 10.0 & 13.5 & 9.6  & 4.3  & 10.7 & \$0.006 \\
Qwen3.5-4B     & 53.3 & 51.2 & 38.9 & 42.7 & 31.8 & 43.6 & \$0.009 \\
Qwen3.5-9B     & 52.4 & 54.7 & 40.6 & 49.8 & 34.5 & 46.4 & \$0.010 \\
Qwen3.5-27B    & 66.9 & 63.3 & 46.8 & 59.4 & 47.1 & 56.7 & \$0.11 \\
\midrule
Gemini-2.5-Flash       & 66.3          & 63.0          & 47.4          & 54.0          & 47.6          & 55.7 & \$0.44 \\
Gemini-3.1-Flash-Lite  & 66.8          & 64.2          & 43.6          & 57.6          & 48.9          & 56.2 & \$0.06 \\
Gemini-3-Flash         & \textbf{69.4} & \textbf{65.0} & \textbf{48.8} & \textbf{61.0} & \textbf{53.9} & 59.6 & \$0.38 \\
\bottomrule
\end{tabular}
\end{table}

\paragraph{Search LLM and dollar cost.}
Table~\ref{tab:open_source_llms} reports nDCG@10 on the five StackExchange subsets across the Qwen3.5 open-weight family (0.8B--27B, non-thinking) and three Gemini models, alongside an estimated dollar cost per query (input/output token rates from Google AI Studio and DeepInfra; token counts measured directly from \algoname~runs). Two findings stand out. First, scaling is the dominant lever: quality climbs steeply from Qwen3.5-2B (10.7) to Qwen3.5-4B (43.6) and saturates around Qwen3.5-27B (56.7), which slightly exceeds Gemini-2.5-Flash (55.7) at $\sim$4$\times$ lower cost (\$0.11 vs.\ \$0.44 per query). Second, on the proprietary side, Gemini-3.1-Flash-Lite hits the best accuracy/cost trade-off: 56.2 nDCG@10 at \$0.06/query -- comparable to Qwen3.5-27B at half the cost and within 2.6 points of Gemini-3-Flash (58.8 at \$0.38). Gemini-3-Flash remains the accuracy ceiling and is what we use for the main BRIGHT results.

\paragraph{Dynamic-corpus subsets.}
\label{sec:dynamic_corpus}
On the three $^*$-marked subsets (LeetCode, AoPS, TheoremQA-Q), where $>$10K documents are excluded per query, \algoname~trails the best ensembles by 9--13 points. Our tree summaries are computed once over the full corpus, so subtrees near excluded leaves become stale and mislead the search. This is atypical of production retrieval, where corpora rarely change per-query, but it points at a real limitation of any static hierarchical index. \algoname\textsuperscript{++} partly closes the gap on these subsets by leaning on the BM25/BGE side of the ensemble for excluded-aware retrieval. A qualitative failure case is in Appendix~\ref{app:fail_analysis}; Appendix~\ref{app:dynamic_insertion} shows that a search-guided document-insertion mechanism can adapt the index without a full rebuild ($+$2.3 nDCG@10 over a static tree on TheoremQA-T).

\subsection{Generalisation Beyond Reasoning-Intensive Retrieval}
\label{sec:beir}

\begin{wraptable}[9]{r}{0.55\linewidth}
\vspace{-12pt}
\centering
\small
\caption{nDCG@10 on BEIR datasets. \algoname~scales to a 2.68M-document corpus on NQ.}
\label{tab:beir_results}
\begin{tabular}{lcccc}
\toprule
\textbf{Dataset} & \textbf{Corpus} & \textbf{BGE-large} & \textbf{BGE-rer.} & \textbf{\algoname} \\
\midrule
SciFact & 5K    & 74.6 & 74.1 & \textbf{75.8} \\
SciDocs & 25K   & \textbf{22.6} & 17.0 & 22.4 \\
NQ      & 2.68M & 55.0 & \textbf{69.0} & 66.8 \\
\bottomrule
\end{tabular}
\vspace{-10pt}
\end{wraptable}

To test whether LLM-guided hierarchical search remains competitive beyond reasoning-intensive retrieval, we evaluate \algoname~on three traditional BEIR~\citep{thakur2021beir} datasets (Table~\ref{tab:beir_results}): SciFact (5K), SciDocs (25K), and NQ (\textbf{2.68M}). \algoname~is competitive on all three -- 75.8 on SciFact (above both BGE-large and BGE-reranker), 22.4 on SciDocs (between the two), and 66.8 on NQ (well above BGE-large 55.0 and within 2.2 of BGE-reranker 69.0) -- and the NQ result additionally shows the approach scales to million-document corpora. That said, on these workloads \algoname~does not add enough value over standard retrieve-then-rerank to justify its substantially higher per-query cost; the regime where it is worth the compute is reasoning-intensive retrieval, where the cheap retriever structurally fails.

\subsection{Analysis and Ablations}
\label{sec:ablations}

Unless stated otherwise, ablations use Gemini-3-Flash as the search LLM and report nDCG@10 on Biology (or StackExchange average where indicated).

\begin{wraptable}[10]{r}{0.55\linewidth}
\vspace{-12pt}
\centering
\small
\caption{Top-down LLM-guided construction vs.\ bottom-up RAPTOR-style embedding clustering.}
\label{tab:ablation_tree}
\begin{tabular}{lcccc}
\toprule
& \multicolumn{2}{c}{\textbf{Biology}} & \multicolumn{2}{c}{\textbf{TheoremQA-T}} \\
\cmidrule(lr){2-3} \cmidrule(lr){4-5}
\textbf{Construction} & nDCG@10 & R@100 & nDCG@10 & R@100 \\
\midrule
Bottom-Up & 67.6 & 89.5 & 33.5 & 64.6 \\
Top-Down  & \textbf{69.4} & \textbf{91.3} & \textbf{46.0} & \textbf{76.2} \\
\bottomrule
\end{tabular}
\vspace{-10pt}
\end{wraptable}

\paragraph{Tree construction.}
Table~\ref{tab:ablation_tree} compares bottom-up RAPTOR-style construction (embed + spectral cluster) against our top-down LLM-guided construction under identical structural constraints ($M$, depth). Top-down wins on both evaluated datasets, with a 2-point gain on Biology and a much larger 12-point gain on TheoremQA-Theorems. We attribute the TheoremQA gap to embedding-based clustering's well-known weakness on dense mathematical notation: theorems that share surface symbols but address unrelated concepts cluster together, producing semantically incoherent subtrees that the search LLM then cannot navigate.

\begin{figure}[!t]
    \centering
    \begin{subfigure}[b]{0.35\linewidth}
        \label{fig:calibration_ablation}
        \centering
        \includegraphics[width=\linewidth]{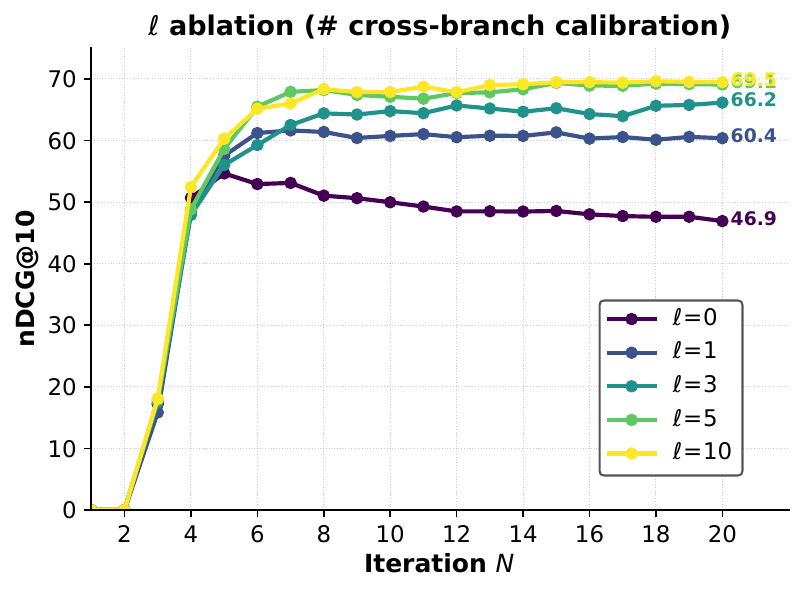}
    \end{subfigure}
    \hspace{0.1\linewidth}
    \begin{subfigure}[b]{0.35\linewidth}
        \label{fig:beam_iterations}
        \centering
        \includegraphics[width=\linewidth]{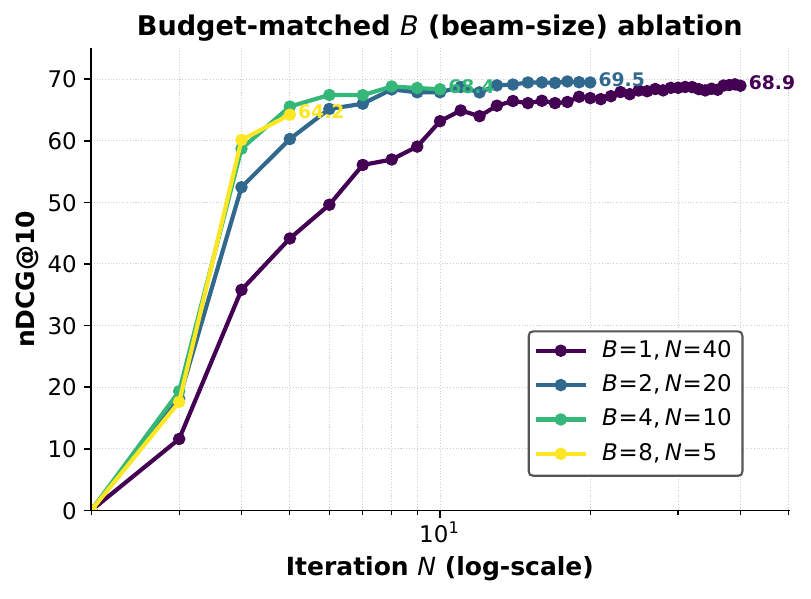}
    \end{subfigure}
    \caption{Hyperparameter ablations on Biology (Gemini-3-Flash). (left) Cross-branch calibration is essential. (right) Under a fixed budget $B\times N{=}40$, depth beats breadth. Trends match on Gemini-2.5-Flash (Appendix~\ref{app:ablation_g25}).}
    \label{fig:hyperparameter_ablations}
\end{figure}

\paragraph{Cross-branch calibration.}
Figure~\ref{fig:hyperparameter_ablations} (left) sweeps the number of cross-branch calibration nodes $\ell$. Without calibration ($\ell{=}0$), Gemini-3-Flash plateaus at 46.9 nDCG@10 -- a 22.6-point gap from the $\ell{=}10$ asymptote of 69.5. Performance improves monotonically and saturates around $\ell{=}5$--$10$ (69.1, 69.5). This supports that slate-dependent score bias is intrinsic to listwise prompting. A small $\ell$ of $5$ already captures most of the gain.

\paragraph{Beam size vs.\ iterations.}
Figure~\ref{fig:hyperparameter_ablations} (right) fixes the search budget at $B\times N{=}40$ and varies the split. Depth dominates breadth: $B{=}1, N{=}40$ and $B{=}2, N{=}20$ reach 68.9 and 69.5; $B{=}4, N{=}10$ drops to 68.4; $B{=}8, N{=}5$ collapses to 64.2. The ordering is identical on Gemini-2.5-Flash (Appendix~\ref{app:ablation_g25}). This justifies our default $B{=}2$ with moderate $N$: it preserves intra-iteration parallelism (two LLM calls in flight per step) while still giving the search enough rounds to commit to a path, refit calibration, and re-rank.

\begin{wraptable}[9]{r}{0.45\linewidth}
\vspace{-12pt}
\centering
\small
\caption{Ablation on \algoname's core traversal components (StackExchange avg., nDCG@10).}
\label{tab:ablation_core}
\begin{tabular}{lc}
\toprule
\textbf{Configuration} & \textbf{Avg.} \\
\midrule
\textbf{\algoname~(Full)}            & \textbf{52.00} \\
\midrule
$-$ No Score Calibration             & 49.56 \\
$-$ No Path Relevance ($\alpha{=}0$) & 48.62 \\
$-$ No Reasoning                     & 49.33 \\
\bottomrule
\end{tabular}
\vspace{-10pt}
\end{wraptable}

\paragraph{Traversal components.}
Table~\ref{tab:ablation_core} ablates each traversal component on StackExchange average. This ablation uses Gemini-2.5-Flash rather than our default Gemini-3-Flash because the Gemini-3-Flash API does not expose a way to disable internal reasoning, so the ``no reasoning'' row (\texttt{thinking\_budget}$=$0) cannot be run with it. All three components contribute meaningfully; path-relevance aggregation is the largest single contributor ($+3.4$ nDCG@10 vs.\ $\alpha{=}0$), followed by explicit reasoning ($+2.7$ vs.\ \texttt{thinking\_budget}$=$0) and score calibration ($+2.4$ vs.\ raw slate scores). The ordering matches the intuition that each component addresses a distinct failure mode: path-relevance gives the search a global signal that individual slate scores; reasoning lets the LLM carefully evaluate the options; calibration separates node-intrinsic relevance from slate-specific bias so cross-branch comparisons remain meaningful.

\section{Related Work}

\paragraph{LLMs in Retrieve-then-Rerank Pipelines.}
The dominant IR paradigm pairs a cheap retriever with an LLM reranker (pointwise or listwise)~\citep{reddy2024first,Sun2023-oa,zhu2023llmir}, with performance bottlenecked by the initial retrieval stage~\citep{rathee2025guiding}. LLMs are also increasingly used as backbones for dense embedding models~\citep{luo2024large, lee2025gemini}, though adapting their autoregressive pre-training to representation learning is awkward.

\paragraph{Generative and Long-Context Retrieval.}
Alternative paradigms include Generative Retrieval (e.g.\ DSI~\citep{tay2022transformer, li2024learning}), which maps queries directly to document IDs but struggles to scale and update~\citep{pradeep2023does}, and Long-Context Retrieval, which loads the entire corpus into the LLM's context~\citep{lee2024can,gupta2025blockrank} but is computationally infeasible at moderate scale. \algoname~offers a middle ground via a semantic hierarchy the LLM navigates efficiently.

\paragraph{Hierarchical Indices.}
Hierarchical structures are long-standing for efficient search in large output spaces -- hierarchical softmax~\citep{morin2005hierarchical}, tree-based extreme classification~\citep{prabhu2014fastxml, yu2022pecos, gupta2022elias}, and HNSW for ANN~\citep{malkov2018efficientrobustapproximatenearest} -- but those hierarchies are geometric, not semantic, and traversal is a vector dot-product. Closer to our setting, RAPTOR~\citep{sarthi2024raptor} builds a semantic hierarchy by bottom-up clustering and summarisation, and EHI~\citep{gupta2024ehi} learns a hierarchical index jointly with a dense encoder; both still traverse via embedding similarity, inheriting the noise of the underlying embedding model at every level. \algoname~differs by using an LLM as an \emph{active traversal agent} that reads node content and reasons at each step.

\paragraph{Adaptive Corpus-Graph Retrieval.}
GAR~\citep{macavaney2022gar}, QUAM~\citep{rathee2024quam}, and SlideGAR~\citep{rathee2025guiding} alternate between an initial-pool reranker and graph-neighbour expansion over a precomputed corpus graph within a fixed reranker-call budget; RGS~\citep{xu2025rgs} performs a similar greedy search over an ANN proximity graph. These are the flat-graph counterparts of our hierarchical traversal. \algoname~differs in that the search structure is an LLM-built \emph{semantic} hierarchy rather than an embedding-built proximity graph that is optimised for nearest-vector search and inherits the limitations of the underlying embeddings.

\paragraph{LLM-based Clustering.}
A related line uses LLMs to organise documents into coherent clusters or topics. TopicGPT~\citep{pham2024topicgpt} prompts an LLM to discover topics top-down, and~\citet{viswanathan2024llmclustering} show that LLMs cluster more reliably than embedding $k$-means on semantically nuanced corpora. Our top-down construction (Section~\ref{sec:construction}) extends this into a recursive hierarchical setting.

\paragraph{Reasoning as Pre-processing and Agentic IR.}
Query expansion (QE)~\citep{wang2023query2doc,gao2023precise} brings LLM reasoning to bear before retrieval but leaves the core search mechanism unchanged, often producing complex multi-component pipelines~\citep{long2025diver,Shao2025-qk}. Agentic IR~\citep{jin2025search, zhang2024agentic, he2025pasa} treats retrieval as a multi-step process where the LLM probes an opaque search tool with generated queries -- effectively guessing keywords to find a local neighbourhood of documents, which is brittle when the right answer is not anticipated in the LLM's parametric knowledge. Graph-RAG~\citep{edge2024local, zhang2025survey} uses LLMs to reason over pre-structured knowledge graphs but with a limited retrieval role. \algoname~differs in that the LLM \emph{is} the search mechanism: the semantic tree provides essential scaffolding that constrains the agent's action space to make global search tractable, while the agent's reasoning enables intelligent traversal -- a setting close in spirit to LLM-guided tree search for reasoning~\citep{yao2023tot,hao2023rap}.

\FloatBarrier

\section{Conclusion}

The standard retrieve-then-rerank pipeline is bottlenecked by the cheap retriever, a problem that becomes acute on reasoning-intensive workloads where it cannot place the right documents in its top-$k$ in the first place. Existing fixes push the LLM upstream of retrieval via query rewriting or agentic loops, but the LLM still interacts with the corpus only indirectly through the same noisy retriever. \algoname~brings the LLM \emph{inside} the retrieval step: it walks an LLM-navigable semantic hierarchy directly, with no embedding model in the loop at search time. To reliably make this paradigm work with off-the-shelf LLMs, we propose two ingredients: a top-down LLM-guided construction that produces an index the LLM can reliably navigate, and a calibrated, path-aggregated traversal that mitigates the slate-dependence and locality of raw LLM scores. With a single zero-shot LLM, \algoname~matches the best four-component fine-tuned ensembles on BRIGHT; a three-source ensemble \algoname\textsuperscript{++} sets a new state-of-the-art at 49.1 nDCG@10. The approach is LLM-agnostic and runs on open-weight backbones at $\sim$4$\times$ lower per-query cost. Because each search step is an LLM call, \algoname~is best suited to quality-over-latency applications (deep research, legal QA, technical QA); natural next steps include search-guided index updates for dynamic corpora and distilling the traversal into smaller models.

\clearpage

\bibliography{references}

\clearpage

\appendix

\begin{figure*}[!thbp]
    \centering
    \includegraphics[width=0.9\linewidth]{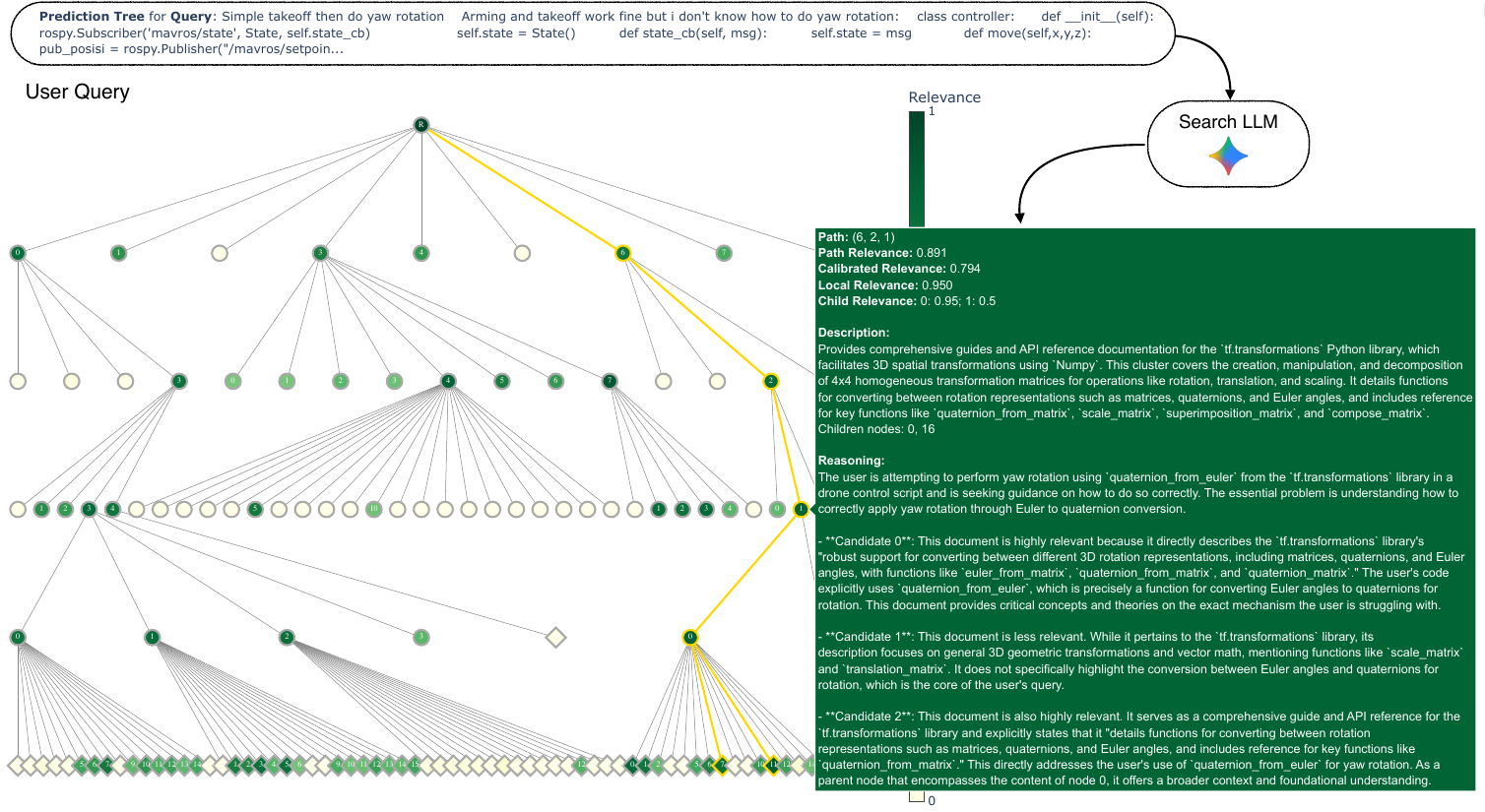}
    \vspace{-10pt}
    \caption{An illustration of the search process of \algoname~for a real query from the BRIGHT benchmark. The color of each node corresponds to its computed path relevance; the highlighted yellow path shows the path to ground-truth documents. The search LLM makes a step-by-step decision at each internal node to determine which branch to explore next. The expanded callout provides a "glass box" view into one such decision, detailing the LLM's explicit reasoning process as it scores the child nodes.}
    \label{fig:sample_search}
    \vspace{-6pt}
\end{figure*}

\section{Limitations and Future Work}
\label{app:limitations}

Our work introduces a promising framework for hierarchical search, but it also presents several avenues for future research. One of the limitation of our current approach is the use of a \textbf{static semantic tree}. Future work could explore methods for efficient, localized updates to the tree's summaries, allowing the hierarchy to adapt to a changing corpus without the need for a full reconstruction.

Second, the \textbf{offline tree construction process}, while a one-time cost, can be computationally intensive for extremely large corpora due to the repeated use of LLMs for clustering and summarization. Research into more efficient construction methods, perhaps by combining traditional clustering for the lower levels with LLM-based summarization for only the top, most abstract layers, could further improve scalability.

While our greedy, best-first traversal is effective in a zero-shot setting, the entire process could be framed as a reinforcement learning problem, where the search LLM is an agent trained to optimize a policy for navigating the tree to maximize retrieval rewards. We believe that exploring these directions will further establish hierarchical, LLM-driven navigation as a powerful new paradigm in information retrieval.

\section{Implementation Details}
\label{app:implementation}

\subsection{Hyperparameters}
\label{app:hyperparams}

This section provides a detailed list of all hyperparameters and implementation choices used in our experiments to ensure full reproducibility.

\subsubsection{Offline Tree Construction}
\label{app:tree_construction}
\begin{itemize}
    \item \textbf{Maximum Branching Factor ($M$):} We set the maximum number of children for any node to $M=10-20$.
    \item \textbf{Embedding Model ($\mathcal{E}$):} For bottom-up tree implementation we use \texttt{gecko}~\citep{lee2024gecko} embeddings to generate vector representations for the clustering steps.
    \item \textbf{Clustering Algorithm ($\mathcal{C}$):} For bottom-up tree implementation uses an iterative spectral clustering~\citep{ng2001spectral} algorithm to partition nodes into at most $M$ clusters at each level of the hierarchy.
    \item \textbf{Summarization LLM:} We use \texttt{Gemini-2.5-flash} for all summarization tasks (both for internal nodes in the bottom-up method and for the multi-level document summaries in the top-down method). The exact prompt template used is detailed in Appendix~\ref{app:prompts}.
    \item \textbf{Top-Down Summary Levels:} For the top-down method, we generate 5 levels of hierarchical summaries for each document.
\end{itemize}

\paragraph{StackExchange Tree Construction.}
For datasets where documents are passages from a smaller set of source articles (the StackExchange sub-datasets in BRIGHT), we leverage this inherent structure. We form initial clusters by grouping all passages belonging to the same source document. If any resulting cluster contains more than $M$ passages, we further subdivide it by grouping passages based on their location proximity within the source document, continuing until all sub-clusters satisfy the branching factor constraint. This approach produces more coherent initial clusters than embedding-based grouping alone, since passages from the same source document are often semantically related.

\subsubsection{Online Traversal}
\begin{itemize}
    \item \textbf{Search LLM ($\mathcal{L}$):} We use \texttt{Gemini-2.5-flash} as the search agent that performs the listwise scoring. The prompt structure is provided in Appendix~\ref{app:prompts}.
    \item \textbf{Number of Iterations ($N$):} We run the search for $N=20$ iterations for all main experiments.
    \item \textbf{Beam Size ($B$):} We use a beam size of $B=2$ for parallel node expansion in each iteration.
    \item \textbf{Path Relevance Momentum ($\alpha$):} The smoothing factor for the path relevance score is set to $\alpha=0.5$.
    \item \textbf{Calibration Nodes ($l$):} We augment each leaf slate with $\ell=10$ cross-branch leaf nodes for calibration, based on our ablation study.
    \item \textbf{Reasoning Budget:} The default ``thinking budget'' for the LLM's reasoning step is set to \texttt{-1}, meaning the model gets to decide how long it wants to thin.
    \item \textbf{MLE Solver:} The latent scores are updated after each batch of slate evaluations. The MSE loss is minimized using the Adam optimizer with a learning rate of \texttt{$10^{-2}$} for \texttt{100} steps.
\end{itemize}

\paragraph{Usage of LLMs} During the preparation of this manuscript, LLM were used as a collaborative writing assistant to aid with drafting, refining prose for clarity and conciseness, and structuring arguments; all core ideas, experiments, and analyses were conducted by the authors.

\subsection{Dataset Details}
\label{app:datasets}

All experiments are conducted on the BRIGHT benchmark~\citep{Su2024-xk}, a comprehensive collection of 12 datasets designed to evaluate reasoning-intensive retrieval. A summary of the statistics for each subset is provided in Table~\ref{tab:dataset_stats}.

\begin{table*}[h]
\caption{Statistics for the 12 subsets of the BRIGHT benchmark used in our experiments.}
\label{tab:dataset_stats}
\begin{center}
\begin{small}
\resizebox{0.6\textwidth}{!}{%
\begin{tabular}{l c c c}
\toprule
\textbf{Dataset Subset} & \textbf{\# Queries} & \textbf{Corpus Size ($\mathcal{D}$)} & \textbf{Avg. Doc Length} \\
\midrule
\multicolumn{4}{l}{\textit{StackExchange}} \\
\midrule
\quad Biology & 103 & 57,359 & 83.6 \\
\quad Earth Science & 116 & 121,249 & 132.6 \\
\quad Economics & 103 & 50,220 & 120.2 \\
\quad Psychology & 101 & 52,835 & 118.2 \\
\quad Robotics & 101 & 61,961 & 121.0 \\
\quad Stack Overflow & 117 & 107,081 & 704.7 \\
\quad Sustainable Living & 108 & 60,792 & 107.9 \\
\midrule
\multicolumn{4}{l}{\textit{Coding}} \\
\midrule
\quad LeetCode & 142 & 413,932 & 482.6 \\
\quad Pony & 112 & 7,894 & 98.3 \\
\midrule
\multicolumn{4}{l}{\textit{Math}} \\
\midrule
\quad AoPS & 111 & 188,002 & 250.5 \\
\quad TheoremQA-Q & 194 & 188,002 & 250.5 \\
\quad TheoremQA-T & 76 & 23,839 & 354.8 \\
\bottomrule
\end{tabular}%
}
\end{small}
\end{center}
\vskip -0.1in
\end{table*}

The datasets exhibit two key characteristics relevant to our work. First, the StackExchange subsets are composed of passages derived from longer source documents. We leverage this structure for our metadata-based initial clustering in the bottom-up tree construction method. Second, the Coding and Theorem-based datasets (excluding Pony and TheoremQA Theorems) utilize a \textbf{query-dependent corpus}, where a unique list of documents (often >10k) must be excluded from the search space for each query. This feature, discussed in our main results analysis, poses a unique challenge for static index structures like our semantic tree.

\subsection{Tree Construction}
\label{app:clustering}
\subsubsection{Bottom-up}
The bottom-up tree construction algorithms are defined in Algorithm~\ref{alg:construction} and Algorithm~\ref{alg:create_nodes}.
\begin{algorithm}[!htbp]
\caption{Bottom-Up Tree Construction}
\label{alg:construction}
\begin{algorithmic}[1]
\STATE \textbf{Parameters:} Corpus $D$, $\mathcal{E}$, $\mathcal{C}$, Summarize LLM, $M$, Optional InitialClusters
\STATE \textbf{Initialize:} $V_L \leftarrow \{ \text{Node}(d) \mid d \in D \}$, $V \leftarrow V_L$, $E \leftarrow \emptyset$

\IF{InitialClusters is provided}
    \STATE $V_{\text{current}} \leftarrow \text{CreateNodesFromClusters}(V_L, \text{InitialClusters}, V, E)$
\ELSE
    \STATE Embeddings $\leftarrow \{ \mathcal{E}(\phi(v)) : v \in V_L \}$
    \STATE Clusters $\leftarrow \mathcal{C}(\text{Embeddings})$
    \STATE $V_{\text{current}} \leftarrow \text{CreateNodesFromClusters}(V_L, \text{Clusters}, V, E)$
\ENDIF

\WHILE{$|\text{V}_{\text{current}}| > M$}
    \STATE \COMMENT{Summarize the current layer before clustering}
    \FORALL{$v$ in $V_{\text{current}}$}
        \STATE $\phi(v) \leftarrow \text{Summarize}(\{ \phi(c) \mid c \in C(v) \})$
    \ENDFOR
    
    \STATE $V_{\text{next\_layer}} \leftarrow \emptyset$
    \STATE Embeddings $\leftarrow \{ \mathcal{E}(\phi(v)) : v \in V_{\text{current}} \}$
    \STATE Clusters $\leftarrow \mathcal{C}(\text{Embeddings})$
    \STATE $V_{\text{next\_layer}} \leftarrow \text{CreateNodesFromClusters}(V_{\text{current}}, \text{Clusters}, V, E)$
    \STATE $V_{\text{current}} \leftarrow V_{\text{next\_layer}}$
\ENDWHILE

\STATE $v_{root} \leftarrow \text{NewInternalNode}()$, $\phi(v_{root}) \leftarrow ""$
\STATE $C(v_{root}) \leftarrow V_{\text{current}}$
\STATE $V \leftarrow V \cup \{v_{root}\}$, $E \leftarrow E \cup \{ (v_{root}, c) \mid c \in C(v_{root}) \}$
\STATE \textbf{return} Tree $T = (V, E)$
\end{algorithmic}
\end{algorithm}

\begin{algorithm}[!htbp]
\caption{CreateNodesFromClusters Subroutine}
\label{alg:create_nodes}
\begin{algorithmic}[1]
\STATE \textbf{function} CreateNodesFromClusters($V_{\text{source}}$, Clusters, $V$, $E$)
\STATE \textbf{Input:}
\STATE \quad $V_{\text{source}}$: The set of nodes in the layer to be clustered.
\STATE \quad Clusters: The partition of $V_{\text{source}}$'s embeddings from $\mathcal{C}$.
\STATE \quad $V, E$: The global node and edge sets for the tree (passed by reference).
\STATE \textbf{Initialize:} $V_{\text{new\_layer}} \leftarrow \emptyset$

\FORALL{cluster $K$ in Clusters}
    \STATE $v_{new} \leftarrow \text{NewInternalNode}()$
    \STATE $C(v_{new}) \leftarrow \{ v \in V_{\text{source}} \mid v \in K \}$
    \STATE $V \leftarrow V \cup \{v_{new}\}$
    \STATE $E \leftarrow E \cup \{ (v_{new}, c) \mid c \in C(v_{new}) \}$
    \STATE $V_{\text{new\_layer}} \leftarrow V_{\text{new\_layer}} \cup \{v_{new}\}$
\ENDFOR

\STATE \textbf{return} $V_{\text{new\_layer}}$
\end{algorithmic}
\end{algorithm}

\subsubsection{Top-down}
\label{app:top_down}
The top-down tree construction algorithm is presented in Algorithm~\ref{alg:top_down_construction} (Section~\ref{sec:construction}). Here we detail the two subroutines it relies on.

The \textbf{SelectSummaryLevel} function implements a heuristic to find the optimal summary granularity for a given set of leaf nodes. It begins with the most abstract summary level ($i=1$) and iteratively checks the number of unique summaries, selecting the first level $i$ where the count of unique summaries is sufficient for meaningful clustering (e.g., greater than $M$) while remaining under a maximum token limit for the LLM context.

The \textbf{ClusterLLM} function is realized via a structured prompt (see Figure~\ref{fig:cluster_llm_prompt}). The LLM is provided with the list of unique summaries and tasked with grouping them into $M$ coherent conceptual clusters. The prompt instructs the model to first generate a short, descriptive title for each of the $M$ clusters, and then to output a mapping from each input summary to one of these cluster titles. The final output is a structured object containing the $M$ topic descriptions (which become the $\phi(v)$ for the new nodes) and the mapping.
\begin{algorithm}[!htbp]
\caption{Top-Down Divisive Tree Construction}
\label{alg:top_down_construction}
\begin{algorithmic}[1]
\STATE \textbf{Parameters:} Corpus $D$, Summarize LLM, Cluster LLM, Max branching factor $M$
\STATE \textbf{Initialize:}
\STATE For each document $d_l \in D$, generate multi-level summaries $\{\phi(v_l)^i\}_{i=1}^5$.
\STATE $V_L \leftarrow \{ \text{Node}(d) \mid d \in D \}$, $V \leftarrow V_L$
\STATE $v_{root} \leftarrow \text{NewInternalNode}()$, $C(v_{root}) \leftarrow V_L$
\STATE $V \leftarrow V \cup \{v_{root}\}$, $E \leftarrow \{ (v_{root}, c) \mid c \in V_L \}$
\STATE PartitionQueue $\leftarrow$ new Queue()
\IF{$|V_L| > M$}
    \STATE PartitionQueue.enqueue($v_{root}$)
\ENDIF

\WHILE{PartitionQueue is not empty}
    \STATE $v \leftarrow$ PartitionQueue.dequeue()
    \STATE LeafDescendants $\leftarrow$ GetLeafDescendants($v, T$)
    \STATE $i \leftarrow$ SelectSummaryLevel(LeafDescendants)
    \STATE UniqueSummaries $\leftarrow \text{unique}(\{ \phi(c)^i \mid c \in \text{LeafDescendants} \})$
    
    \STATE TopicDescs, Mapping $\leftarrow \text{ClusterLLM}(\text{UniqueSummaries}, M)$
    
    \STATE NewChildren $\leftarrow \emptyset$
    \FOR{$j = 1$ \TO $M$}
        \STATE $v'_j \leftarrow \text{NewInternalNode}()$, $\phi(v'_j) \leftarrow \text{TopicDescs}[j]$
        \STATE $V \leftarrow V \cup \{v'_j\}$, NewChildren $\leftarrow$ NewChildren $\cup \{v'_j\}$
    \ENDFOR
    
    \STATE ReassignChildren(LeafDescendants, Mapping, NewChildren, T)
    \STATE $E \leftarrow E \setminus \{ (v, c) \mid c \in C(v) \}$ \COMMENT{Disconnect old children}
    \STATE $C(v) \leftarrow$ NewChildren
    \STATE $E \leftarrow E \cup \{ (v, c) \mid c \in \text{NewChildren} \}$ \COMMENT{Connect new children}
    
    \FORALL{$v'_{j}$ in NewChildren}
        \IF{$|C(v'_{j})| > M$}
            \STATE PartitionQueue.enqueue($v'_{j}$)
        \ENDIF
    \ENDFOR
\ENDWHILE

\STATE \textbf{return} Tree $T = (V, E)$
\end{algorithmic}
\end{algorithm}

\subsection{Tree Traversal}
\label{app:traversal_algo}
The full pseudocode for the online \algoname~tree search described in Section~\ref{sec:traversal} is given in Algorithm~\ref{alg:traversal}.
\begin{algorithm}[!htbp]
    \caption{LLM-guided Hierarchical Search}
    \label{alg:traversal}
    \begin{algorithmic}[1]
    \STATE \textbf{Parameters:} $q, T, \mathcal{L}, B, N, K, \alpha$
    \STATE \textbf{Initialize:}
    \STATE Frontier $F \leftarrow$ new MaxPriorityQueue(), $\text{Pred} \leftarrow \emptyset$
    \STATE $\text{ScoreHistory} \leftarrow \emptyset$, $\text{LatentScores} \leftarrow \emptyset$
    \STATE $\hat{p}_{rel}(v_{root}) \leftarrow 1.0$, $F.\text{push}(v_{root}, \hat{p}_{rel}(v_{root}))$

    \FOR{$i=1$ \TO $N$}
        \STATE Beam $\leftarrow$ Extract top $B$ nodes from $F$

        \FORALL{$v$ in Beam}
            \STATE Slate $\leftarrow C(v) + Aug(v)$
            \STATE LocalScores $[s_{v'}]_{v' \in \text{Slate}} \leftarrow \mathcal{L}(q, [\phi(v')]_{v' \in \text{Slate}})$
            \STATE Add $\{(\text{slate\_id}_i, v', s_{v'}) \mid v' \in \text{Slate}\}$ to $\text{ScoreHistory}$
        \ENDFOR

        \STATE $\text{LatentScores} \leftarrow \text{SolveMLE}(\text{ScoreHistory})$ \COMMENT{Minimize MSE to find all $\hat{s}_v$}

        \FORALL{$v$ in Beam that were just expanded}
            \FORALL{$v'$ in Slate}
                \STATE $\hat{s}_{v'} \leftarrow \text{LatentScores}[v']$
                \STATE $\hat{p}_{rel}(v') \leftarrow \alpha \cdot \hat{p}_{rel}(\text{parent}(v')) + (1-\alpha) \cdot \hat{s}_{v'}$
            \ENDFOR
            \FORALL{$v'$ in $C(v)$}
                \IF{$v'$ is a leaf node}
                    \STATE Add $v'$ to $\text{Pred}$
                \ELSE
                    \STATE $F.\text{push}(v', \hat{p}_{rel}(v'))$
                \ENDIF
            \ENDFOR
        \ENDFOR
    \ENDFOR

    \STATE \textbf{return} Top-$K$ nodes from $\text{Pred}$ sorted by $\hat{p}_{rel}$
    \end{algorithmic}
\end{algorithm}

\subsection{\algoname\textsuperscript{++}: Per-Component Ensemble Analysis}
\label{app:ensemble_ablation}

The main paper reports \algoname\textsuperscript{++} (Section~\ref{sec:results}), a low-cost ensemble that augments \algoname~with BM25 and the BGE-Reasoner-Embed dense retriever. For each query, we min-max normalise the per-document scores from each system to $[0,1]$ and take a weighted sum with weights $0.6$ (\algoname), $0.2$ (BM25), and $0.2$ (BGE-Reasoner-Embed). BM25 and BGE-Reasoner-Embed are both run over the GPT-4 reasoning-rewritten queries released with BRIGHT, following the standard evaluation protocol on the benchmark.

\begin{table*}[!htbp]
    \caption{Per-component decomposition of \algoname\textsuperscript{++} (nDCG@10); fusion scheme described in Section~\ref{sec:setup_exp}. \algoname~uses Gemini-2.5-Flash; BM25 and BGE-Reasoner-Embed run on the GPT-4 rewritten queries. $^*$~denotes subsets with a dynamic corpus.}
    \label{tab:ensemble_ablation}
    \begin{center}
    \begin{small}
    \resizebox{\textwidth}{!}{%
    \begin{tabular}{l|c|ccccccc|c|cc|c|ccc|c}
    \toprule
    \textbf{Configuration} & \textbf{SE} & \textbf{Bio.} & \textbf{Earth.} & \textbf{Econ.} & \textbf{Psy.} & \textbf{Rob.} & \textbf{Stack.} & \textbf{Sus.} & \textbf{Cod.} & \textbf{Leet.$^*$} & \textbf{Pony} & \textbf{Thm} & \textbf{AoPS$^*$} & \textbf{ThQ.$^*$} & \textbf{ThT.} & \textbf{Avg.} \\
    & \textbf{Avg.} & & & & & & & & \textbf{Avg.} & & & \textbf{Avg.} & & & & \\
    \midrule
    BM25 (rewritten query)              & 36.4 & 58.3 & 55.2 & 23.4 & 38.3 & 24.8 & 26.7 & 27.9 & 25.3 & 23.8 & 26.8 & 20.1 & 5.6  & 29.5 & 25.1 & 30.5 \\
    BGE-Reasoner-Embed (rewritten query) & 43.4 & 56.7 & 59.1 & 33.2 & 45.7 & 32.3 & 39.3 & 37.4 & 34.0 & 26.8 & 41.3 & 37.2 & 15.6 & 45.0 & 51.1 & 40.3 \\
    \algoname~(Gemini-2.5-Flash)         & 52.0 & 66.3 & 63.0 & 47.4 & 54.0 & 47.6 & 37.6 & 48.2 & 26.9 & 19.9 & 34.0 & 32.6 & 12.0 & 38.0 & 47.9 & 43.0 \\
    \midrule
    \algoname\textsuperscript{++} (Gemini-2.5-Flash) & \textbf{53.6} & \textbf{71.4} & \textbf{66.9} & 46.5 & \textbf{55.3} & 45.4 & \textbf{43.7} & 45.9 & \textbf{36.7} & \textbf{28.5} & \textbf{44.8} & \textbf{36.7} & \textbf{14.2} & \textbf{43.6} & \textbf{52.4} & \textbf{46.5} \\
    \bottomrule
    \end{tabular}%
    }
    \end{small}
    \end{center}
\end{table*}

Table~\ref{tab:ensemble_ablation} decomposes \algoname\textsuperscript{++} (Gemini-2.5-Flash) into its three components. \algoname~alone is the strongest single retriever on average (43.0 nDCG@10), more than 2 points ahead of BGE-Reasoner-Embed (40.3) and 12 points ahead of BM25 over the reasoning-rewritten query (30.5). The ensemble closes a further 3.5-point gap, reaching 46.5. The gains over single \algoname~are concentrated in two places: (i) the dynamic-corpus subsets (LeetCode $19.9 \to 28.5$, AoPS $12.0 \to 14.2$, TheoremQA-Q $38.0 \to 43.6$, TheoremQA-T $47.9 \to 52.4$), where stale tree summaries hurt the static \algoname~index and the two retrievers contribute orthogonal signal, and (ii) the StackExchange subsets dominated by document-level topic match (Biology $66.3 \to 71.4$, Earth Science $63.0 \to 66.9$), where the dense retriever is already strong and the ensemble accumulates a few additional correct retrievals on top of \algoname's high baseline. On subsets where \algoname~is already well above both retrievers -- Economics ($47.4$ vs.\ $33.2$/$23.4$), Robotics ($47.6$ vs.\ $32.3$/$24.8$), Sustainable Living ($48.2$ vs.\ $37.4$/$27.9$) -- ensembling provides little additional gain.

\subsection{Hyperparameter Ablations with Gemini-2.5-Flash}
\label{app:ablation_g25}

\begin{figure}[!htbp]
    \centering
    \begin{subfigure}[b]{0.42\linewidth}
        \centering
        \includegraphics[width=\linewidth]{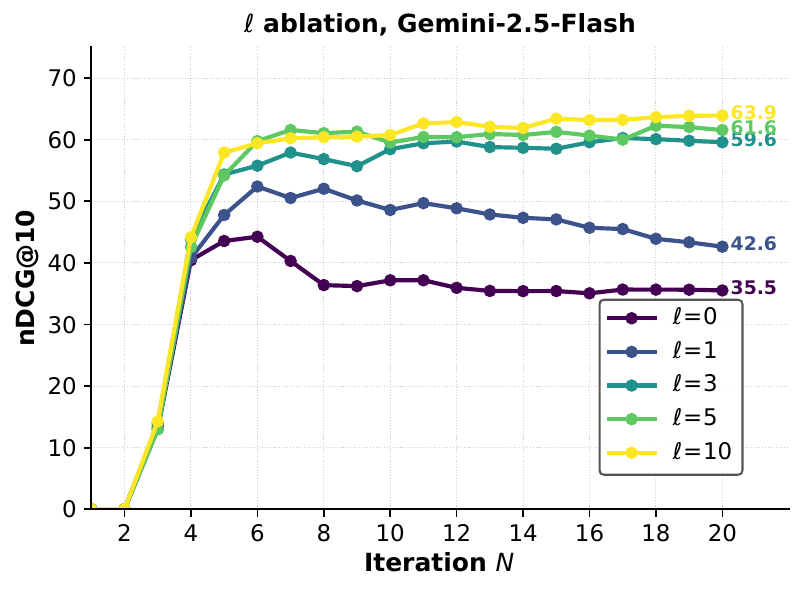}
        \caption{Cross-branch calibration nodes $\ell$.}
    \end{subfigure}
    \hspace{0.04\linewidth}
    \begin{subfigure}[b]{0.42\linewidth}
        \centering
        \includegraphics[width=\linewidth]{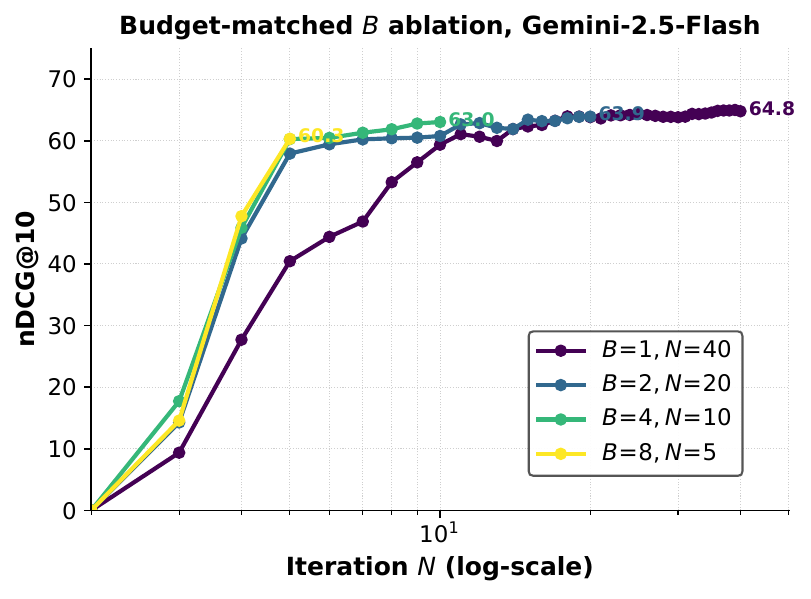}
        \caption{Beam size $B$ vs.\ iterations $N$ under $B{\times}N{=}40$.}
    \end{subfigure}
    \caption{Hyperparameter ablations on Biology with Gemini-2.5-Flash. Trends mirror the Gemini-3-Flash version in Figure~\ref{fig:hyperparameter_ablations} -- calibration is essential and depth beats breadth -- at uniformly lower absolute nDCG@10.}
    \label{fig:hyperparameter_ablations_g25}
\end{figure}

For completeness, Figure~\ref{fig:hyperparameter_ablations_g25} reports the same hyperparameter sweeps as Figure~\ref{fig:hyperparameter_ablations} (Section~\ref{sec:ablations}) but with Gemini-2.5-Flash as the search LLM. Both trends carry over: nDCG@10 improves monotonically with $\ell$ (calibration is essential) and depth beats breadth under a fixed budget. The absolute scores are 4--10 points lower than with Gemini-3-Flash, but the relative ordering of configurations is identical.

\subsection{Cost Per Query Breakdown for Table~\ref{tab:open_source_llms}}
\label{app:cost_breakdown}

The cost-per-query column of Table~\ref{tab:open_source_llms} is computed as $\bar{T}_{\text{in}} \cdot p_{\text{in}} + \bar{T}_{\text{out}} \cdot p_{\text{out}}$, where $\bar{T}_{\text{in}}, \bar{T}_{\text{out}}$ are the average input and output token counts per query measured directly from \algoname~runs and $p_{\text{in}}, p_{\text{out}}$ are per-token rates from the providers. Per-token rates (\$/1M, in/out): Google AI Studio -- Gemini-2.5-Flash $0.30 / 2.50$, Gemini-3-Flash $0.50 / 3.00$, Gemini-3.1-Flash-Lite $0.25 / 1.50$; DeepInfra (Qwen3.5) -- 0.8B $0.01 / 0.05$, 2B $0.02 / 0.10$, 4B $0.03 / 0.15$, 9B $0.04 / 0.15$, 27B $0.26 / 2.40$. Measured per-query token counts: Qwen (non-thinking, family avg) $\bar{T}_{\text{in}}{\approx}149$K, $\bar{T}_{\text{out}}{\approx}27$K; Gemini-2.5-Flash $168$K / $155$K; Gemini-3-Flash $164$K / $98$K; Gemini-3.1-Flash-Lite $153$K / $14$K.

\subsection{Retrieval Coverage on StackExchange}
\label{app:recall_coverage}

\begin{figure}[!htbp]
    \centering
    \includegraphics[width=0.7\linewidth]{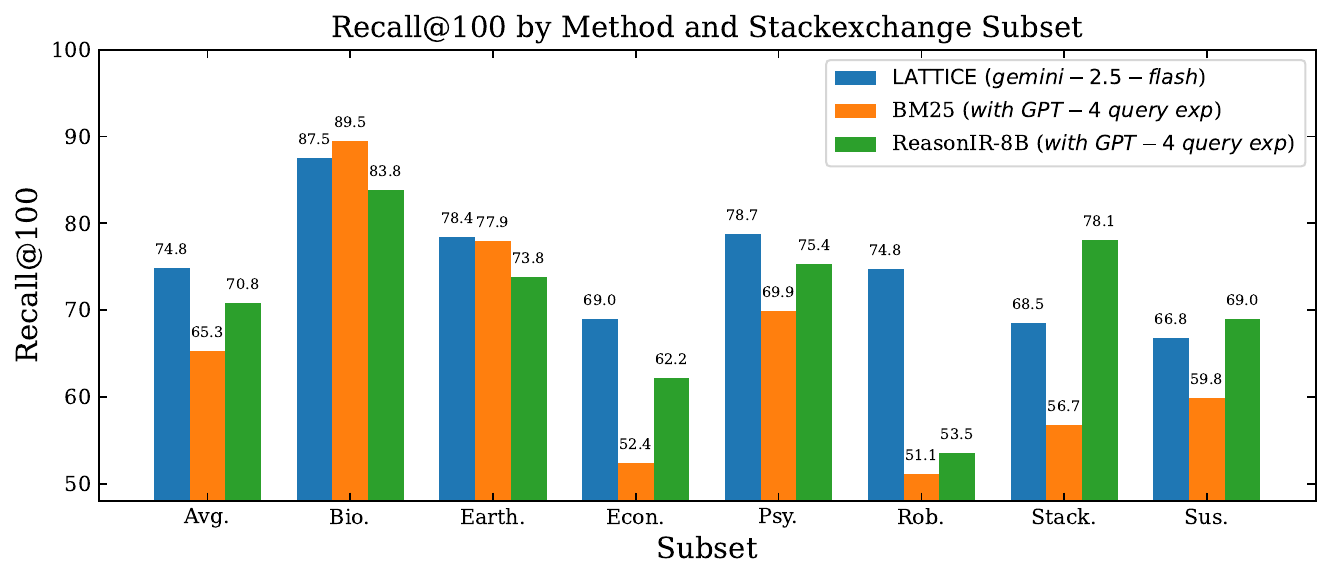}
    \caption{Recall@100 on the seven StackExchange subsets of BRIGHT. \algoname~achieves 74.8\% average recall, outperforming BM25 by $+9.5$ pp and the fine-tuned ReasonIR-8B by $+4.0$ pp.}
    \label{fig:recall_results}
\end{figure}

Beyond top-10 ranking, \algoname~surfaces a substantially better candidate set than the strongest retrieval-only baselines. Figure~\ref{fig:recall_results} reports Recall@100 on the seven StackExchange subsets of BRIGHT: \algoname~(Gemini-2.5-Flash) achieves 74.8\% average recall -- $+9.5$pp over BM25 over the GPT-4 rewritten query, and $+4.0$pp over the fine-tuned ReasonIR-8B. The largest coverage gains are on the multi-step-reasoning subsets, consistent with the ranking pattern reported in the main paper: when the relevance signal requires reasoning, walking the tree exposes documents that flat first-stage retrievers miss entirely.

\subsection{Search-and-Insert Index Updates}
\label{app:dynamic_insertion}

A practical concern for any hierarchical index is adaptability to corpus changes: each insertion or deletion can in principle invalidate cluster summaries along the affected root-to-leaf path. We explore a lightweight \emph{search-and-insert} update mechanism that side-steps a full rebuild. Given a new document, we run the online search of Section~\ref{sec:traversal} on the existing tree using the document text itself as the query, and attach the document as a leaf of the highest-scoring leaf cluster returned by the search. The cluster's $\phi(\cdot)$ summary is left unchanged.

\begin{table}[!t]
\centering
\small
\caption{Dynamic insertion on TheoremQA-T. Documents placed via search-guided insertion outperform a static tree built with the full corpus.}
\label{tab:dynamic_insertion}
\begin{tabular}{lcc}
\toprule
\textbf{Method} & \textbf{nDCG@10} & \textbf{R@100} \\
\midrule
Static Construction (full corpus)     & 47.4 & 73.9 \\
Dynamic Insertion (search-guided)     & \textbf{49.7} & \textbf{81.1} \\
\bottomrule
\end{tabular}
\end{table}

We evaluate this mechanism on TheoremQA-Theorems by withholding all ground-truth documents during tree construction, building a tree over the remaining corpus, and then inserting the withheld documents via the search-guided procedure above. Table~\ref{tab:dynamic_insertion} compares retrieval quality against the static baseline -- the same tree built once over the full corpus, including the gold documents.

The dynamically updated tree (49.7 nDCG@10, 81.1 Recall@100) \emph{outperforms} the static tree (47.4 nDCG@10, 73.9 Recall@100). Although this result is on a single dataset and we do not claim it as a general phenomenon, it is consistent with a familiar intuition from HNSW-style graph construction~\citep{malkov2018efficientrobustapproximatenearest}: when document placement is made by the same procedure that later retrieves the document, the resulting index is structurally well-aligned with the search algorithm that will traverse it.


\section{Subjective Analysis}
\label{app:analysis}
\subsection{Sample scoring response from LLM}
To provide a more intuitive understanding of our method, Figure~\ref{fig:sample_search} presents a qualitative case study of the search process for a real query from the BRIGHT benchmark. The user query is a code snippet asking about ``yaw rotation,'' a complex 3D graphics problem. The figure visualizes the semantic tree and the traversal path taken by \algoname~(highlighted in yellow) to successfully locate a relevant document deep within the hierarchy.

The expanded callout provides a "glass box" view into the search LLM's reasoning at a critical decision point. The LLM's generated \textbf{Reasoning} explicitly connects the user's query to the node's topic, noting that the user is ``attempting to perform yaw rotation using quaternion\_from\_euler.'' It then performs a detailed, comparative evaluation of the children nodes. It correctly identifies Candidate 1 as highly relevant because it discusses ``support for converting between different 3D rotation representations, including matrices, quaternions, and Euler angles,'' which directly addresses the user's problem. This example demonstrates that our method does not rely on shallow semantic similarity; instead, the search is an active process guided by the LLM's deep, step-by-step reasoning about the query in the context of the corpus hierarchy.

\subsection{Search failure on dynamic corpus}
\label{app:fail_analysis}
\begin{figure}
    \centering
    \includegraphics[width=\linewidth]{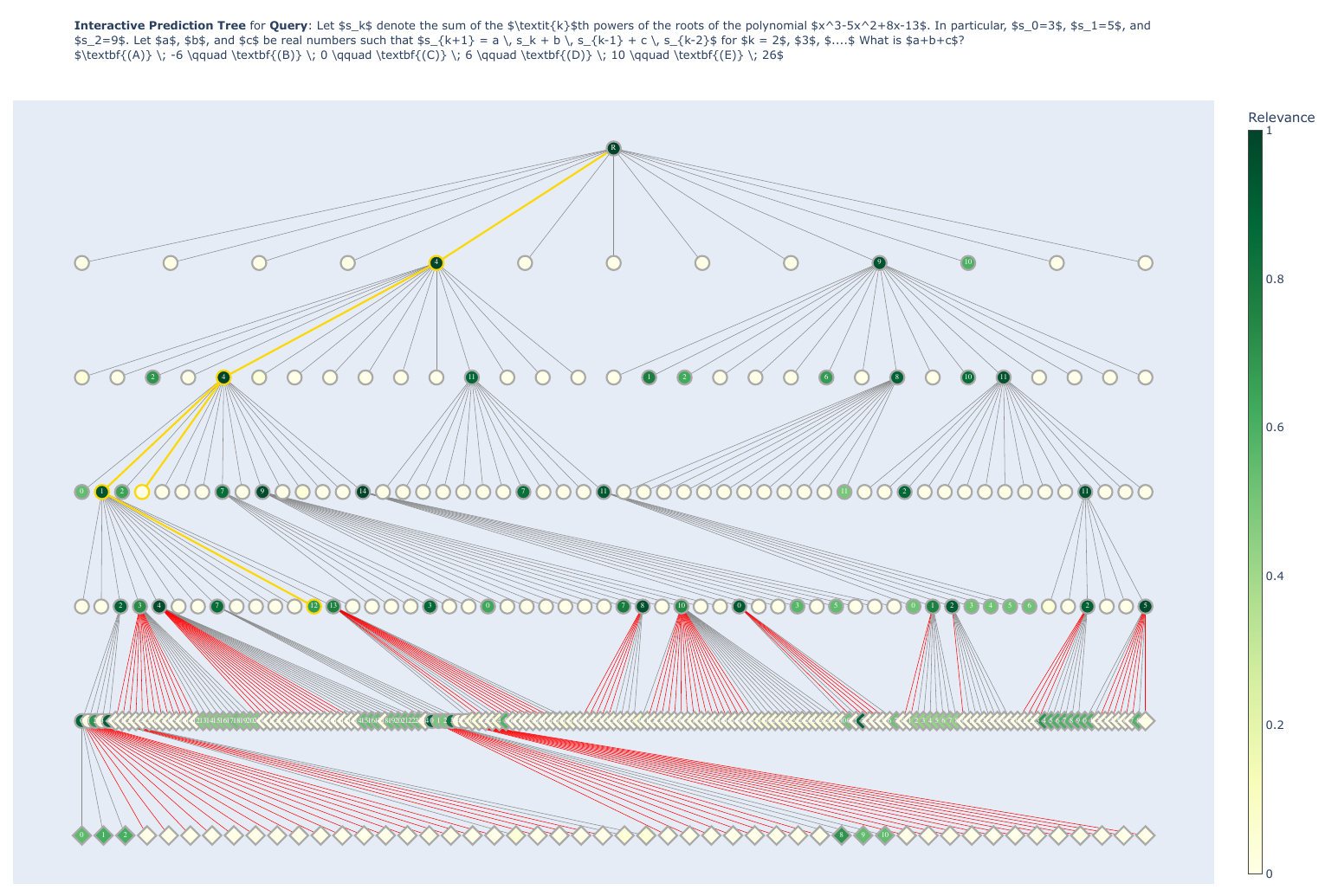}
    \caption{Search failing due to dynamically excluded search corpus, {\color{red}red edges} denote excluded leaf nodes, {\color{yellow}gold edges} denote ground-truth path}
    \label{fig:aops_fail}
\end{figure}

Figure~\ref{fig:aops_fail} provides a qualitative case study of a search failure, visually demonstrating the primary challenge our method faces on datasets with a dynamic corpus. The figure shows the search tree for a random query from the AoPS dataset. Red edges indicate leaf nodes that were dynamically excluded for this specific query, while the yellow path highlights the ideal traversal route to the ground-truth document.

As the figure shows, the search agent correctly follows the ground-truth path for the first two levels. However, it then reaches an internal node whose pre-computed summary is now misleading; the summary was generated based on all of its children, including the large number that have since been pruned from the search space (the red nodes). This inaccurate, stale summary causes the search LLM to make an incorrect judgment, deviating from the correct path and ultimately failing to retrieve the relevant document. This example visually confirms the specific failure mode of a static hierarchical index when faced with a dynamic corpus, reinforcing the quantitative analysis in our main results section.

\section{Prompts}
\label{app:prompts}
\tcbset{
  myanswerbox/.style={
    colframe=darkgray,   
    colback=blue!1!white,
    boxrule=1pt,         
    left=1mm, right=0mm, top=1mm, bottom=1mm, 
  },
}

\lstset{basicstyle=\tiny\ttfamily,breaklines=true, breakindent=0pt}

\begin{figure}[thbp] 
\begin{tcolorbox}[myanswerbox]
\scriptsize{\color{blue}
\begin{lstlisting}[breaklines]
You are an intelligent search agent navigating a hierarchical semantic tree of topics. Your mission is to predict the most promising candidates to find the answer to the user's query using the relevance definition below.

**Relevance Definition:** {relevance_defintion}

---

## USER QUERY

{query}

---

## CANDIDATES

Here are the candidates, each is identified by a unique `node_id` provided at the very start in [] (e.g., [0]).

{child_node_options}

---

## YOUR EVALUATION TASK
1.  First, identify the essential problem in the query.
2.  Think step by step to reason about why each candidate is relevant or irrelevant (based on the relevance definition). Provide this analysis in the `reasoning` field.
3.  Rank these passages based on their relevance to the query. Provide your ranking in the `ranking` field.
4.  Assign a relevance score from 0 to 100 (based on the relevance definition and the ranking). Provide relevances in the `relevance_scores` field.

---

## OUTPUT FORMAT
You must provide your response as a single, clean JSON object. The JSON should have three keys: `reasoning`, `ranking`, and `relevance_scores`.

* `reasoning`: This must be a **string**.
* `ranking`: This must be an **array of integers** representing the order of the candidates.
* `relevance_scores`: This must be an **array of arrays** where each inner array contains [node_id, relevance_score]. For example: [[0, 85], [1, 92], [2, 73]].

---

## YOUR RESPONSE
\end{lstlisting}}
\end{tcolorbox}
\caption{Prompt template used in our experiments for scoring a list of nodes for $\mathcal{L}$. 
}
\label{fig:scoring_prompt}
\vspace{-15pt}
\end{figure}
\tcbset{
  myanswerbox/.style={
    colframe=darkgray,   
    colback=blue!1!white,
    boxrule=1pt,         
    left=1mm, right=0mm, top=1mm, bottom=1mm, 
  },
}

\lstset{basicstyle=\tiny\ttfamily,breaklines=true, breakindent=0pt}

\begin{figure}[thbp] 
\begin{tcolorbox}[myanswerbox]
\scriptsize{\color{blue}
\begin{lstlisting}[breaklines]
You are an expert in information retrieval and keyword generation. Your task is to analyze a provided list of informational passages and generate a hierarchically sorted list of search keywords for each passage, strictly adhering to the 5-level rubric below.

## Keyword Generation Rules (5 Levels):

Level 1: 1-2 Word, Core Subject / Domain (Broadest)
Meaning: The absolute fundamental, overarching subject area or discipline.
Characteristics: Only 1 to 2 word, very high-level (e.g., "Technology", "Science", "History")

Level 2: 3-4 Word, General Topic / Sub-domain
Meaning: Narrows Level 1; the specific major topic or branch within the broader field.
Characteristics: Only 3 to 4 words, still general but more focused

Level 3: 4-6 Word, Key Concepts / Main Themes
Meaning: The central ideas, significant concepts, or primary themes directly discussed.
Characteristics: Only 4 to 6 words, core messages, primary subjects, often main sections

Level 4: 7-10 Word, Very Concise Passage Summary
Meaning: A very short, concise summary of what the entire passage is about. This should encapsulate the essential idea or purpose of the passage.
Characteristics: Only 7 to 10 words

Level 5: 11-20 Word, Concise Passage Summary (Most Specific)
Meaning: A concise summary but more descriptive than level 4 of what the entire passage is about. This should encapsulate the main idea or purpose of the passage.
Characteristics: A single sentence, 11 to 20 words.

### General Keyword Requirements:

- All keywords must be actionable terms or phrases a user would realistically search.
- Ensure comprehensive coverage of the passage's content across all 5 levels.

## Output Format

Your output must be a single JSON object. This object will contain a top-level key: "passages_keywords". The value associated with this key must be a JSON array. Each element in this array will be an object with two keys:
"passage_id": An integer that exactly matches the "id" from the corresponding input passage.
"hierarchical_keywords": A JSON array of strings of length 5. Each string represents a hierarchical level (Level 1 at index 0, Level 2 at index 1, and so on).

## List of Input Passages:

{desc_list}
\end{lstlisting}}
\end{tcolorbox}
\caption{Prompt template used in our experiments for generating multi-level keywords to be used in top-down tree construction.
}
\label{fig:multi_level_keyword_prompt}
\vspace{-15pt}
\end{figure}
\tcbset{
  myanswerbox/.style={
    colframe=darkgray,   
    colback=blue!1!white,
    boxrule=1pt,         
    left=1mm, right=0mm, top=1mm, bottom=1mm, 
  },
}

\lstset{basicstyle=\tiny\ttfamily,breaklines=true, breakindent=0pt}

\begin{figure}[thbp] 
\begin{tcolorbox}[myanswerbox]
\scriptsize{\color{blue}
\begin{lstlisting}[breaklines]
You are an expert data analyst and taxonomist. Your task is to analyze a list of keywords and their associated counts which indicate how many that keyword appears in the corpus.

## Goal
- Group the following keywords into **k** semantically coherent and **well-balanced** (i.e. each cluster should aim to contain similar weighted count) clusters, where k is between [{min_k}, {max_k}]. The primary basis for grouping must be the **topic and meaning** of the keywords.
- Use the provided count as a measure of each keyword's **importance or popularity**. This weight should help you decide which topics are most significant.
- Try to always maximize the number of clusters but **without** sacrificing the quality of the clustering, **quality of clustering is paramount**.

For every cluster, generate:
* A descriptive `cluster_name`.
* An information-dense `cluster_description` summarizing the core themes.
* A list of all input `keywords` that constitute this cluster or apply to this cluster.

## Input Data
Here is the list of keywords and their importance counts:

{keywords_list_with_count}

## Desired Output Format
Your final output must be a single JSON object, with no other text or explanation. The JSON object must have key: "clusters".

{{
  "clusters": [
    {{"name": "Name of Cluster 1", "description": "A very information dense description of the cluster", "keywords": ["keyword 1", "keyword 2", ...] }},
    {{"name": "Name of Cluster 2", "description": "A very information dense description of the cluster", "keywords": ["keyword 3", "keyword 4", ...] }},
    ...
  ],
}}

---

## Your Response
\end{lstlisting}}
\end{tcolorbox}
\caption{Prompt template used for ClusterLLM to be used in top-down tree construction i.e. clustering a given set of keywords into $[M_{min}, M_{max}]$ clusters.
}
\label{fig:cluster_llm_prompt}
\vspace{-15pt}
\end{figure}

\tcbset{
  myanswerbox/.style={
    colframe=darkgray,   
    colback=blue!1!white,
    boxrule=1pt,         
    left=1mm, right=0mm, top=1mm, bottom=1mm, 
  },
}

\lstset{basicstyle=\tiny\ttfamily,breaklines=true, breakindent=0pt}

\begin{figure}[thbp] 
\begin{tcolorbox}[myanswerbox]
\scriptsize{\color{blue}
\begin{lstlisting}[breaklines]
You are an expert AI analyst and summarizer. Your mission is to create a highly informative and "discriminative signpost" for a navigating search agent. This signpost (a summary) must guide the agent to the correct cluster of nodes to answer a user's query.

You will follow a strict, step-by-step cognitive process. You must analyze the children nodes in a target parent node (the "Positive Set").

Prompt ID: {prompt_id} (ignore, this is just for watermarking purposes).

## INPUTS

### POSITIVE SET: Information about the target parent node to be summarized

{positive_set_descriptions}
---

## YOUR TASK & OUTPUT FORMAT

Your entire output must be a single, valid JSON object. Inside this JSON, you will follow the 3-step thinking process outlined below, populating each field as instructed.

### JSON Structure and Instructions:

{{
  "detailed_fingerprints": [
    // For EACH children node in the POSITIVE SET (target parent node), extract a structured object of its key, queryable facts.
    {{
      "one_line_summary": "...", // write a very information dense and very concise one-line summary for the information contained in this node
      "key_entities": ["..."], // List a very few key entities which is central to this node
      "genre_or_category": ["..."], // List a few key genre / categories this node can be classified into
      "name": "...", // Name the node
    }}
  ],
  "common_theme": "...", // Reason deeply what are the common themes between the nodes in the POSITIVE SET
  "summary": "...", // Based on step 1 and step 2, write a very information dense description of the target node, **make sure to include all key entities**.
}}

---

## Your Response
\end{lstlisting}}
\end{tcolorbox}
\caption{Prompt template for generating bottom-up summaries of a group of nodes.
}
\label{fig:bottom_up_summarizer_prompt}
\vspace{-15pt}
\end{figure}

\end{document}